%% file: SPHEREx_arXiv.tex
\documentclass[plain]{aiaa-pretty} % {article}%

\usepackage{bm}
\usepackage{amsmath}
\usepackage{subfigure}
\usepackage{algorithm} %http://myitcv.org.uk/latex/tips_and_tricks.html#answer8
\usepackage{algorithmic}  
\usepackage{textcomp}
\usepackage{gensymb}
\usepackage{footnpag}			      	% make footnote symbols restart on each page
\begin{document}
%
%\AIAAconference{AIAA Journal??}
%
%\AIAAnumber{XX}
\title{All Sky Survey Mission Observing Scenario Strategy}
\author{Sara Spangelo \thanks{Systems Engineer and Researcher, Jet Propulsion Lab, California Institute of Technology, 4800 Oak Grove Drive, Pasadena, CA 91109}
Raj Katti \thanks{Technician of Physics at Caltech Institute of Technology, MC 249-17, 1200 East California Blvd, Pasadena CA 91125 }
Steve Unwin \thanks{Exoplanet Exploration Deputy Program Scientist, Jet Propulsion Lab, California Institute of Technology, 4800 Oak Grove Drive, Pasadena, CA 91109}
Jamie Bock \thanks{Professor of Physics at Caltech Institute of Technology and Senior Research Scientist at the Jet Propulsion Laboratory, 4800 Oak Grove Drive, Pasadena, CA 91109}}
%and Ben Longmier \thanks{Assistant Professor, Aerospace Engineering, University of Michigan, 1320 Beal Ave, Ann Arbor, MI 48109.} }
%\footnote{Timestamp: \timestamptwelve }

%
%
%\vspace{-10mm}
\AIAAabstract{
\input{abstract}
}
%, where all solutions are guaranteed to be feasible because of the constraint-based approach.}
%
%The general framework of dividing and solving the sub-problems in series and the constraint-based scheduling approach is applicable to larger classes of complex scheduling problems.}
%\input{sections/abstract}

\maketitle

\input{body}
\bibliographystyle{unsrt}
\bibliography{FGSN_biblio}
%

%\appendix
%\input{sections/app} % Appendix
%\input{sections/app_time_history} % Appendix
% 
\section*{Acknowledgments}  % This is defined in the IEEEAerospace2008.cls and adds an Acknowledgments section heading
%%%%%%%%%%%%%%%%
%
The authors acknowledge Olivier Dore, Timothy Kock, Kirk Breitenbach, Hemali Vyas, Dustin Crumb, and Anthony Pullen for their contributions.
Part of the research was carried out at the Jet Propulsion Laboratory, California Institute of Technology, under a contract with the National Aeronautics and Space Administration.

%We would also like to acknowledge the CubeSat community for their support in completing the ground station survey.%The text of your acknowledgment goes here. This section is optional.
\end{document}

%% file: abstract.tex
This paper develops a general observing strategy for missions performing all-sky surveys, where a single spacecraft maps the celestial sphere subject to realistic constraints. The strategy is flexible such that targeted observations and variable coverage requirements can be achieved.  This paper focuses on missions operating in Low Earth Orbit, where the thermal and stray-light constraints due to the Sun, Earth, and Moon result in interacting and dynamic constraints.  The approach is applicable to broader mission classes, such as those that operate in different orbits or that survey the Earth. First, the instrument and spacecraft configuration is optimized to enable visibility of the targeted observations throughout the year. Second, a constraint-based high-level strategy is presented for scheduling throughout the year subject to a simplified subset of the constraints. Third, a heuristic-based scheduling algorithm is developed to assign the all-sky observations over short planning horizons. The constraint-based approach guarantees solution feasibility. The approach is applied to the proposed SPHEREx mission, which includes coverage of the North and South Celestial Poles, Galactic plane, and a uniform coverage all-sky survey, and the ability to achieve science requirements demonstrated and visualized. Visualizations demonstrate the how the all-sky survey achieves its objectives.

%% file: body.tex
\section{Introduction}

\subsection{Observing Scenario Overview}

This paper develops an general observing strategy approach for accomplishing an all-sky survey, which could be applied to both mapping the celestial sphere (zenith-pointing) or mapping the Earth (nadir-pointing).
%
% spacecraft mission observing scenario consisting of an all sky survey, which is 
 The approach is flexible such that it is applicable to missions with focused observations and those that may have variable coverage requirements.
%consisting (TBC- make flow) of three distinct surveys focusing on different areas of the celestial sky with different wavelength and coverage requirements.
%
%are the fixed angle between the instrument and spacecraft and where the instrument should point at every point in time.
%
%
The approach focuses on missions in a Low Earth Orbit (LEO), although the constraints can be modified or relaxed and applied to a broader range of orbit scenarios.
% and Earth-avoidance constraints and the instrument has a stray-light Moon-avoidance constraint.
%
% that constrain both the spacecraft configuration and observing scenario.
%
The observing problem is dynamic throughout the year as the orbit evolve relative to the Sun, Moon, celestial sphere, and other potential targets (e.g. Galactic plane).
%, thus the constraints are also dynamic.
%
The problem formulation considers the interacting and dynamic constraints related to thermal and stray-light avoidance relative to the Sun, Earth, and Moon for a mission operating in LEO.
%Because the instrument is fixed relative to the spacecraft, 
These combined constraints limit the zone where the spacecraft and instrument can point, which varies throughout the year.
%is more limiting at some times of the year.
%where some times of the year are more constraining than others.
%
Decisions related to both to the spacecraft configuration and observing strategy, 
% decision variables include both the instrument configuration relative to the spacecraft and the observing scenario schedule.
%
%The spacecraft configuration decisions and strategy 
which must be robust throughout the full year, are addressed.
%and applicable to all times of the year.
%
%The observing scenario is a 
System-level issues related to the spacecraft configuration, the telecommunication system, the attitude control system, and the thermal control system, are considered in the strategy.
\subsection{Literature Review}

\input{lit_review_v2}
\subsection{Paper Overview}

The observing scenario problem constraints and objectives are described in detail in Section \ref{sec:prob_desc}.
Capturing this problem as a single global problem would be complex and may be difficult to generate solutions guaranteed optimal solutions.
%or impossible to generate solutions with guaranteed feasibility or optimality.
%
To overcome this challenge, this paper separates and solves these problems in series, passing a simplified set of constraints between sub-problems.
Thus, although solutions may not be guaranteed optimal, they are guaranteed to be feasible.
First, a feasible instrument configuration is established that satisfies the Sun-avoidance constraint and observability requirements throughout the year.
% in Section \ref{sec:obs_str}.\ref{sec:sc_inst_config}.
%
Second, the configuration and Earth-avoidance (thermal) constraints are combined and yield maximum observation time constraints that depend on the angle of the orbital plane relative to the Sun.
%$\beta$.
%
These first two steps are described in Section \ref{sec:obs_str}.\ref{sec:sc_inst_config}.
Third, a high-level scheduling strategy and a resultant
%for assigning the pointings to achieve the requirements and goals of the three surveys, which includes
%, described in Section \ref{sec:obs_str}.\ref{sec:point_scheduling}.
%
%This establishes the cadence and total number of pointings for the various surveys.
%
%Fourth, 
constraint-based heuristic algorithm is developed in Section \ref{sec:obs_str}.\ref{sec:sched_alg}.
% that assigns the all-sky observations, which compliments the assigned pointings for the other surveys.
%
This constraint-based approach guarantees solution feasibility.
Extensions to the special case focusing on the Galactic plane is described in Section \ref{sec:obs_str}.\ref{sec:gal_case}.
Section \ref{sec:appl_SPHEREx} applies the general approach to the SPHEREx mission, an astrophysics LEO mission which includes three district surveys, including an all-sky survey with specific coverage requirements.
%
%We demonstrate how 
The survey goals are achieved in both idealized and realistic cases, and demonstrated with coverage visualizations.
% shown to satisfy the survey requirements and objectives in Section \ref{sec:obs_str}.\ref{sec:implement_scheduling}.
%
%Fifth, the heuristic-based algorithm is extended to accommodate the Moon-avoidance stray-light constraint in Section \ref{sec:obs_str}.\ref{sec:moon_avoid}
%
%Because the problem is highly constrained, a constraint-based approach is taken such that all resulting solutions are feasible, even if they may not be optimal.
%
The paper contributions and results are summarized and insights into how  the step-wise approach presented in this paper can be applied to other observing scenarios are described in Section \ref{sec:conclu}. 

%TBC- fix so SPHEREx intro'd part way.
%\subsection{Paper Overview}
%
%%Discuss required fraction of pointings and how this varies throughout the year (depending on maximum number of total steps/ pointing).
%
%Describe paper contributions.
%%
%The paper presents a heuristic constraint-based scheduling strategy for the SPHEREx mission for both long (1 year) and short timescales (1 day).
%%
%
%Describe how can be generalized to other scheduling problems.

%\newpage
\section{Problem Description}
\label{sec:prob_desc}

%
%This section describes the definitions, objectives, and constraints of an all sky observing scenario problem.
%
%Vectors, angles, and terms used throughout the paper are defined in Section \ref{sec:prob_desc}.\ref{sec:defs} and the objectives and constraints of the scheduling problem are described in Section \ref{sec:prob_desc}.\ref{sec:obs_overview}.
%, and the resulting spacecraft and instrument configuration is described in Section \ref{sec:prob_desc}
%Section \ref{sec:prob_desc}.\ref{sec:obs_overview} provides and overview of the observing scenario 

%   % \begin{wrapfigure}{r}{0.5\textwidth}
%\begin{figure}[!h]
%     \begin{center}
%         \includegraphics[width=.48\textwidth,angle=0]%{figs/angles_sketch}
%     \end{center}     
%\caption{SPHEREx Angle Definitions \label{fig:angles_sketch} }
%   % \end{wrapfigure}
%\end{figure}
%\subsection{Objectives, Assumptions, and Constraints}
%\label{sec:obs_overview}
%%\subsection{Orbit}

This section describes the generic all-sky scheduling problem, by defining vectors, angles, constraints, and then defining scheduling terms used throughout the paper.
It is assumed that the spacecraft is in a Sun synchronous LEO, as typically selected for all-sky surveys, such as IRAS, Akari, and WISE.
Missions with different orbits may still benefit from the algorithms in Section \ref{sec:obs_str}, but may require alternative formations of the constraints presented in this section (in many cases most of these constraints can be relaxed or ignored for orbits other than LEOs, which tend to be particularly constraining).
In Sun synchronous orbits, every rotation about the Earth provides visibility to all declinations.
By definition, Sun synchronous orbits precess at a rate of approximately one degree a day, thus the instrument can access the entire celestial sphere in approximately six months.
%with a redundancy of one in six months and two in one year.
% throughout results in a precession of approximately one degree 
%

%In summary, 
The overall goal of the scheduling problem is to efficiently observe the celestial sphere.
The approach presented in this paper can be applied to pointed observations distributed across the celestial sphere or the case where the entire celestial sphere must be observed.
%scanned within a time constraint.
%
%This can be stated as maximizing science time.
%, or stated otherwise to maximize science time.
%, which is essentially the same as maximizing the Deep survey redundancy.
%
%This goal is accomplished by minimizing the large and small slew durations.
%
The decision variables are where and when to point the instrument on the celestial sphere as a function of time.
The dynamics include the orbit motion relative to the Earth, Sun, Moon, celestial sphere, and other targets (e.g. Galactic plane).
%
%, see Fig. \ref{fig:gal_long_GPcross}.
%
%, and relative to the Sun and Moon.
%
%The constraints include accomplishing the survey redundancy requirements for the various surveys, and
There are constraints related to  
% Galactic and all-sky surveys described in this section, and 
the thermal and stray-light avoidance from the Earth, Sun, and Moon.
There may also be constraints that the mission must accomplish targeted observations or observe certain areas of the celestial sphere more often.

%\subsection{Definitions}
%\label{sec:defs}

\begin{figure}[!hb]
\begin{center}
   \subfigure[View Orthogonal to Orbital Plane]{
  \includegraphics[width=.48\textwidth,angle=0]{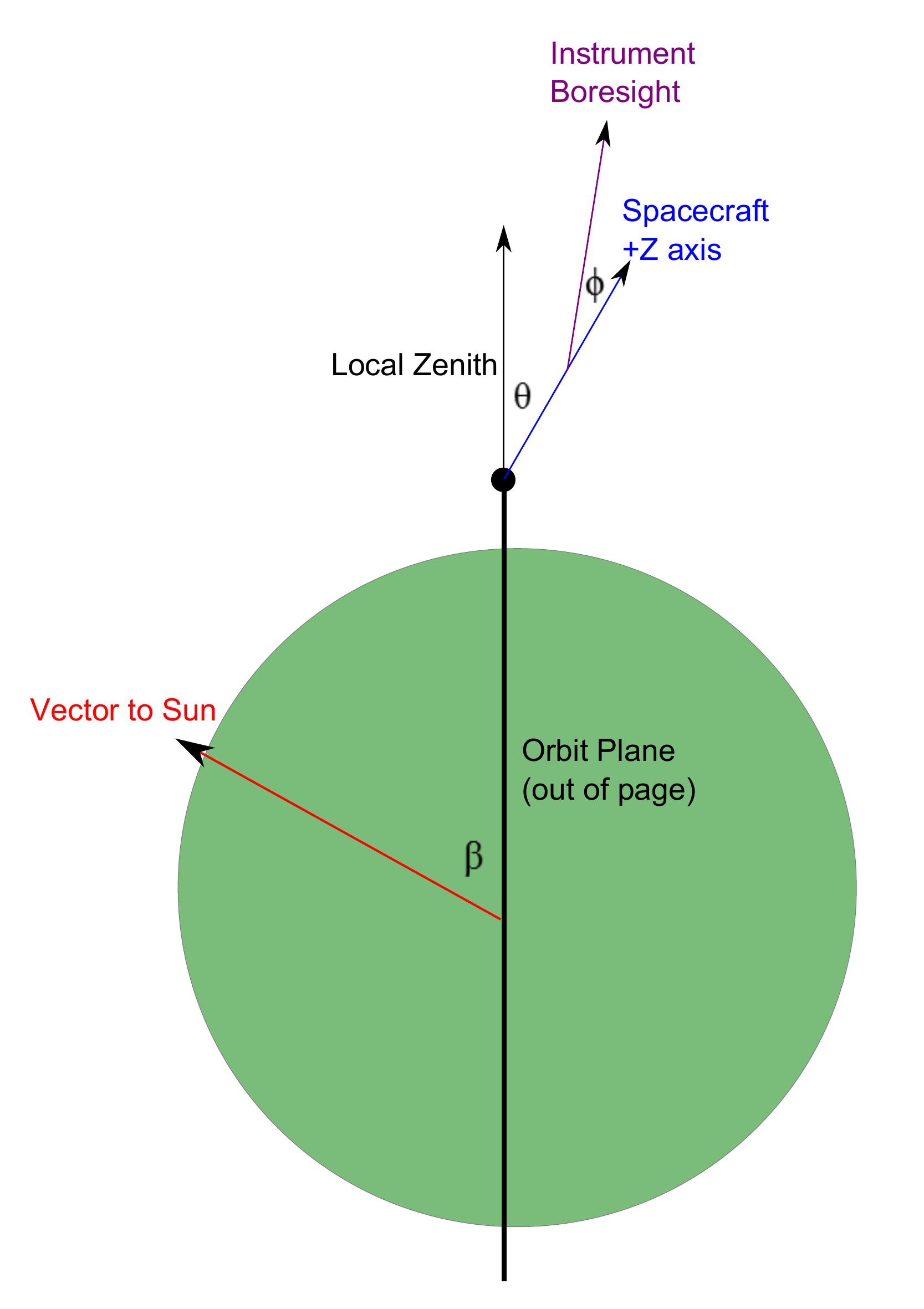}
   \label{fig:angles_sketch}}
   \subfigure[View in Orbital Plane]{
    \includegraphics[width=.48\textwidth,angle=0]{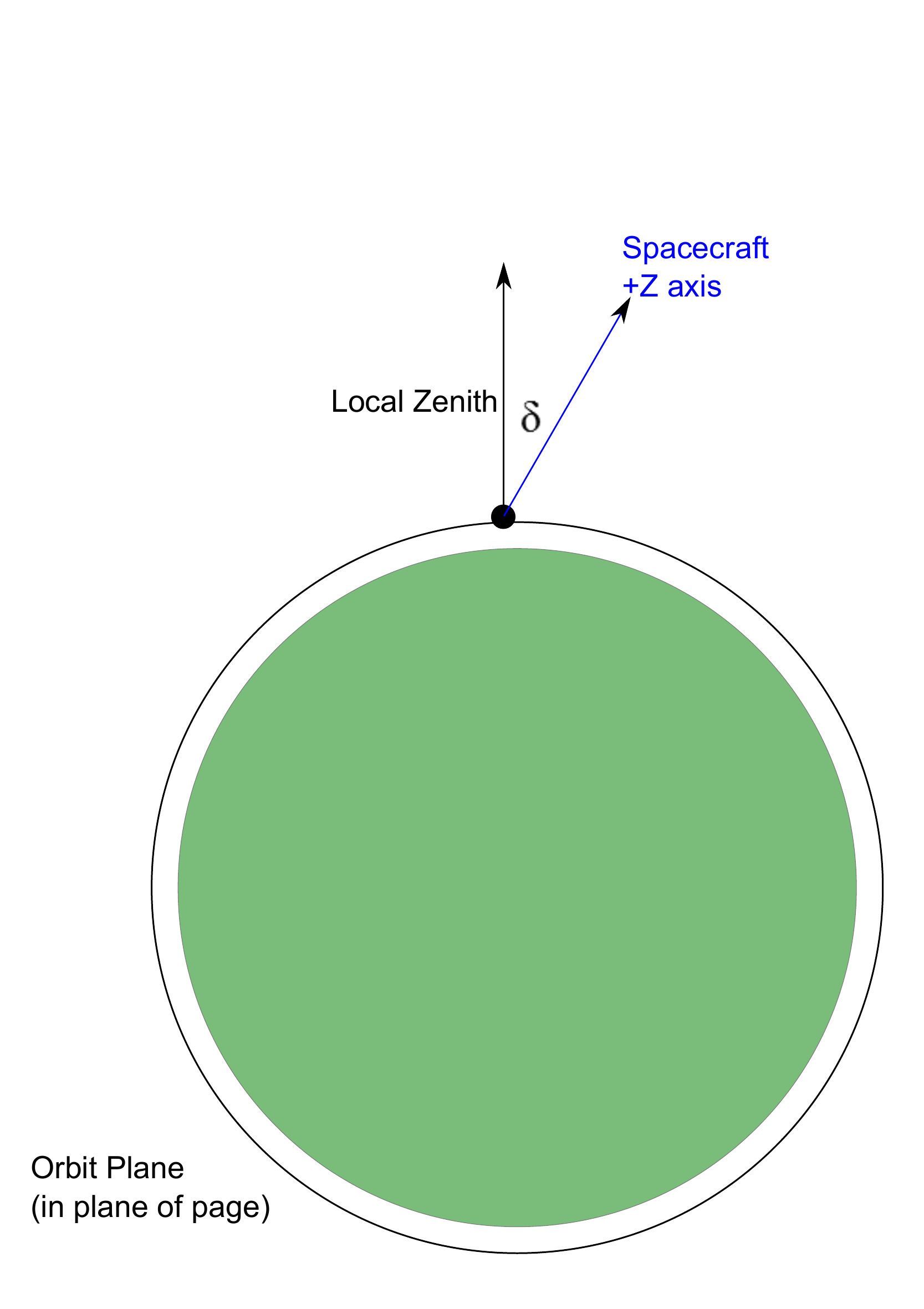}
    \label{fig:angles_sketch_inplane}}
\end{center}
\caption{ Angle Definitions \label{fig:angles_sketch_all}}
\end{figure}

%For this type of mission 
To scan the celestial sphere, the spacecraft points roughly in the zenith direction, and the instrument is fixed at an offset angle, see Fig. \ref{fig:angles_sketch_all}.
%
%\noindent 
Prior to describing the constraints, several vectors and angles are defined:
%vectors used in this paper are defined as:
%we refer to in this paper, which are shown in Figure \ref{fig:angles_sketch}:

\begin{itemize}
\item Spacecraft +Z axis: Symmetry axis of the spacecraft.
%, as defined in \ref{fig:angles_sketch_all}
%
\item Instrument boresight: Vector aligned with the center of the telescope field of view (FOV).
\end{itemize}

%\noindent The angles used in this paper are defined as:
%, and are shown in Figure \ref{fig:angles_sketch}

\begin{itemize}
\item Beta angle ($\beta$): The angle between the orbital plane and vector to the Sun.
\item Tilt angle ($\theta$): The angle between the orbital plane and the spacecraft +Z axis in the orbit cross-track direction.
\item Cant angle ($\phi$): The angle between the spacecraft +Z axis and the instrument boresight.
% in the orbit cross-track direction.
%
\item Nod angle ($\delta$): The angle between the spacecraft +Z axis and the local zenith in the orbital plane.
\end{itemize}

The scheduling problem is also subject to the following spacecraft and instrument constraints (where the angles are 
%defined in Section \ref{sec:prob_desc}.\ref{sec:defs} and
 shown in Fig. \ref{fig:angles_sketch_all}):
%thermal and stray-light 
\begin{itemize}
\item Sun-avoidance criteria: Spacecraft +Z axis cannot be pointed within $\Omega$ of line-of-sight to the Sun due to thermal constraints, 
\vspace{-5mm}
\begin{align}
\label{eq:theta}
\theta \geq \Omega-\beta.
\end{align}
%
%In particular, the spacecraft boresight must point at least $\Omega$ from the line-of-sight to the Sun, 
%This constraint is satisfied when the spacecraft points to the right which is satisfied starting at the Sun Line in Fig. \ref{fig:3D_constraints}.
% shows where this constraint is satisfied.
%
\item Earth-avoidance criteria: Spacecraft +Z axis cannot be pointed more than $\alpha$ from the local zenith  due to thermal constraints.
The spacecraft is rotated both in the cross-track direction by $\theta$ and the in-track 
direction by $\delta$, thus the constraint on the resulting total angle is:
\vspace{-2mm}
\begin{align}
\label{eq:alpha}
\cos(\alpha) = \cos(\theta) \cos(\delta),
\end{align}
%
%where $\alpha$ is the angle between the spacecraft +Z axis and local zenith.
%Note this must account for total angular offsets both in the in-track and cross-track directions.
%
\item Moon-avoidance criteria: Instrument boresight cannot point within $\zeta$ of line-of-sight to the Moon due to stray-light constraints.
%
%In particular, $\zeta$, the angle between the instrument boresight and vector to the Moon, must be less than some value.
% +Z axis and local zenith defined as:
\end{itemize}

\begin{figure}[!h]
 \begin{center}
     \includegraphics[width=.6\textwidth,angle=0]{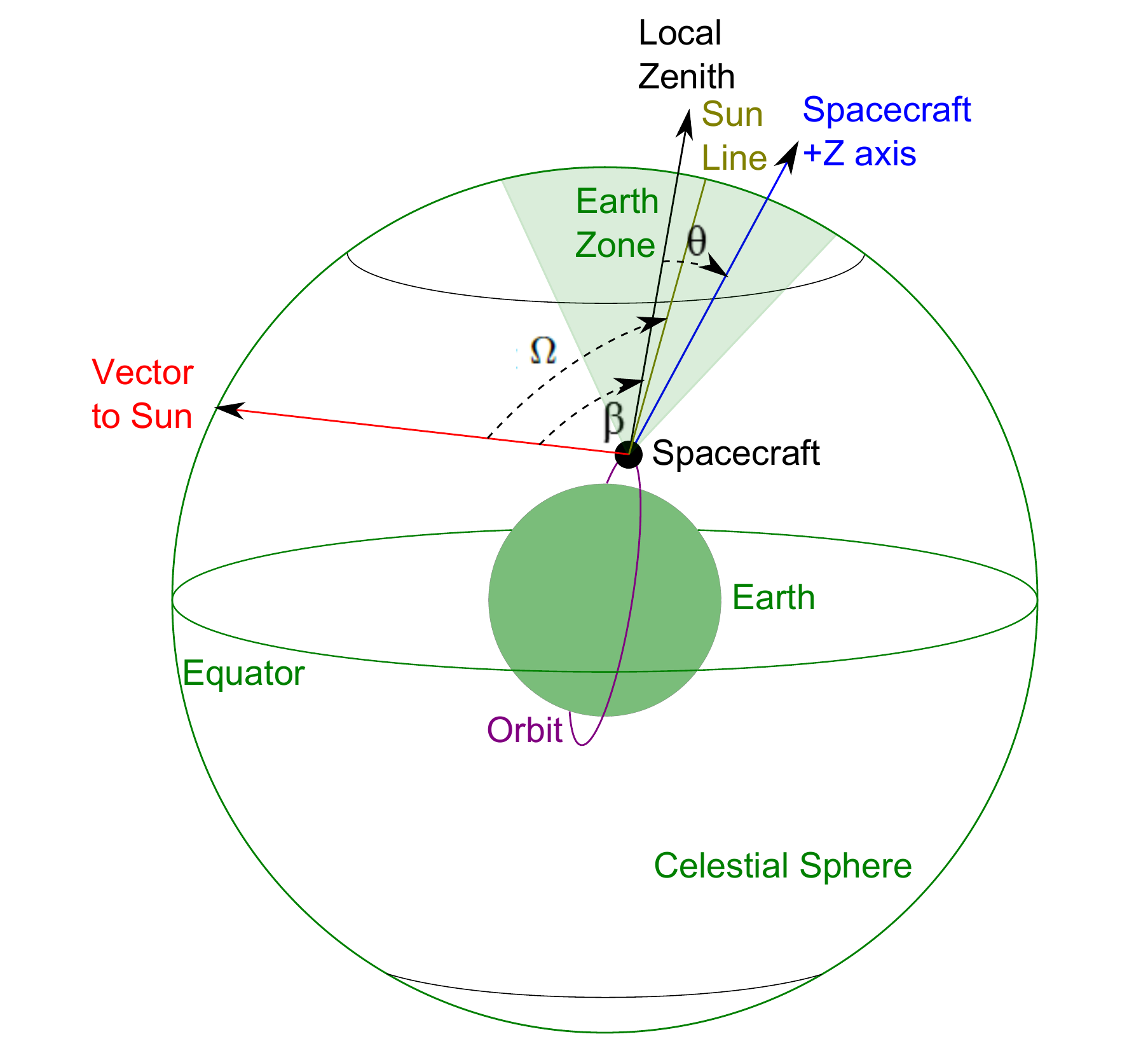}
 \end{center}
 \caption{Constraints on spacecraft orientation.  The spacecraft must point to the right of the Sun Avoidance Line and within the Earth Zone. \label{fig:3D_constraints}}
\end{figure}

These combined constraints result in visibility restrictions for where spacecraft and instrument can point as a function of time.
In particular, the spacecraft boresight must point to the right of the Sun Avoidance Line (where the $\Omega$  Sun-constraint is satisfied) and within the Earth Zone to satisfy the combination of constraints in Fig. \ref{fig:3D_constraints}.

%Include a figure to demonstrate all these angles clearly.

\noindent The scheduling terms used in this paper are defined as:

\begin{itemize}
\item Pointing: Period of time focused on a region of the sky, comprised of steps.
% (with duration on the order of several minutes).
\item Step: Subset of a pointing pointing focused on one target in the sky.
% (with duration on the order of 100-200 seconds).
%
A pointing is composed of a fixed number of steps.
% (e.g. 4-9).
%
%\item Slew: Maneuver to transition between pointings or steps.
\item Large Slew: A maneuver to transition between pointings.
\item Small Slew: A maneuver to transition between steps.
\item Redundancy: Number of times the same region of the sky is observed by the desired part of the instrument FOV appropriate for that survey (i.e. if a region of the sky is observed once by every wavelength, the redundancy is one).
\end{itemize}

Throughout the remainder of the paper the observing scenario is described in terms of the number of pointings per orbit and number of steps per pointing.

%  \begin{figure}[!h]
%   \begin{center}
%       \subfigure[Spacecraft pointing direction constraints ($\beta=60^{\circ}$) ]{
%      \includegraphics[width=.45\textwidth,angle=0]{figs/3D_constraints_beta90}
%       \label{fig:3D_constraints_beta90}}
%%
%       \subfigure[Spacecraft pointing direction constraints ($\beta=90^{\circ}$) ]{
%        \includegraphics[width=.45\textwidth,angle=0]{figs/3D_constraints_beta60}
%        \label{fig:3D_constraints_beta60}}
%   \end{center}
%   \caption{Constraints on where spacecraft can point, where it must point to the right of the Sun Line and within the Earth Zone.
%   %
%%   The $\beta=60^{\circ}$ case is far more constrained, with a cone of only  
%\label{fig:3D_constraints}}
%  \end{figure}

\section{Observing Scenario Strategy}
\label{sec:obs_str}

The general all-sky observing strategy is to point the spacecraft roughly in the zenith direction in the orbit plane throughout every orbit.
% for two reasons.
%
Both the ascending and descending portions of the orbit are utilized equally for observing.
First, this enables mapping of the entire celestial sphere because the orbital plane naturally precesses at a rate of one degree per day, enabling coverage of the entire celestial sphere in six months.
% and 
%Second, the instrument will generally target a specific area on the sky for long durations to maximize the science time (by minimizing the number of pointings and thus number of slews).
%%
%Furthermore, pointing in the roughly zenith direction 
Second, this maximizes the time before the Earth-avoidance constraints will be violated by minimizing the number of pointings and thus number of slews (i.e. the instrument can point at a target in the sky for a long time).

% which is required to achieve the redundancy requirements of the all-sky survey.
%
%This section introduces the observing scenario approach.
%

%Provide section overview?
%Section \ref{sec:obs_str}.\ref{sec:sc_inst_config} describes on the spacecraft and instrument configuration to achieve the scenario objectives, and Section 

\subsection{Spacecraft and Instrument Configuration}
\label{sec:sc_inst_config}

To satisfy the Sun-avoidance constraint, the general strategy is to tilt the spacecraft according to $\theta = \Omega-\beta$ as in Eq. \ref{eq:theta}.
% (see Fig. \ref{fig:angles_sketch}).
% to always satisfy Sun-avoidance criteria.
%
To satisfy the Earth-avoidance criteria in Eq. \ref{eq:alpha}, the nod in the in cross-track direction, $\delta$, is constrained,
\begin{align}
\label{eq:delta}
\delta_{max} \leq \cos^{-1}\left(\frac{\cos(\alpha)}{\cos(\theta)}\right)
\end{align}
This restricts the time the instrument can focus at a single target, where the maximum time is a function of two times the half angle and the orbital period (where the following equation accounts for the conversion from degrees to radians),
\begin{align}
\label{eq:Tmax}
T_{max} = \frac{\delta_{max} P}{ \pi},
\end{align} %
where $P$ is the orbital period and $\delta_{max}$ is given in radians.
The maximum pointing times vary as a function of $\beta$ throughout the year. %, as in Fig. \ref{fig:Tmax_yr}.
This places dynamic constraints on the problem, where $T_{max}$ varies throughout the year, and as a result, the minimum number of pointings per orbit is higher for lower $\beta$ values.
An example is given in Fig. \ref{fig:Tmax_yr} for the SPHEREx mission described in the next section.
%at times of the year the  spacecraft can dwell on a target for 19 minutes, while 
%Describe limitations for $\beta=60^o$ case, major being that the maximum time is 9 minutes, as in Fig. \ref{fig:Tmax_yr}.
%

%\subsection{General Strategy}
%\label{sec:obs_gen}

%Major differences when $\beta=60^o$, for example deep surveys have only 3 185 sec pointings to fit into 9 minutes.
%%
%NCP/SCP scheduled every second opportunity (instead of every opportunity).
%%
%Galactic plane pointings consist of only 4 pointings each, and thus occur nearly every orbt.
%%
%Reference example in Fig. \ref{fig:pointings_beta60}.

%Describe how we arrived at the cant angle (to view NCP/SCP year-round).

\subsection{Scheduling Algorithm}
\label{sec:sched_alg}

The general scheduling approach is to prioritize the most constrained pointings and steps and schedule these first, while satisfying the dynamic constraints such as the $\beta$ and position of the targets (e.g. Galactic Plane).
Second, the all-sky algorithm generates a schedule that maps the complete celestial with the desired redundancy in the time of interest while satisfying all constraints.
% to realistic 
% crossing throughout the scheduling time horizon (which is orbit-dependent).
%
%The approach described in this section ignores the Moon-avoidance.
%, discussed in Section \ref{sec:obs_str}.\ref{sec:moon_avoid}. 
%
The approach is constraint-based, so by construction yields feasible solutions.
%
%While good solutions are found, they
Solutions are not guaranteed to be globally optimal; however they do provide good initial guesses for further optimization.
%

%First, the Galactic Plane pointings are scheduled according to the required cadence and strategy for covering the entire plane throughout the year.
%to when it is observable throughout the year on the desired cadence.
% and distribution throughout the year to obtain the required fraction of steps per orbit.
%
First, the targeted pointings are scheduled according to the desired cadence and when the target regions are accessible, which depends on the time of the year.
%, as described in Section \ref{sec:obs_str}.\ref{sec:sc_inst_config}.
%
The cadence for the these observations is selected to achieve the required total number of steps per time that satisfies coverage requirements and is feasible with the rest of the survey.
%
%Deep Survey observations are not feasible on every orbit because of the need to cover high-declination all-sky areas.
%%
%%per year for these two surveys, 
%For example see the fractions in Table \ref{tab:schedule_beta90} for the $\beta=90^{\circ}$ case.
%
%The scheduling of the Galactic Plane and Deep surveys could be performed in any order as they do not directly interfere with one another.
%
Second, the all-sky pointings are scheduled, which is more complex and described in greater detail next.

    \begin{figure}[!h]
     \begin{center}
         \includegraphics[width=.6\textwidth,angle=0]{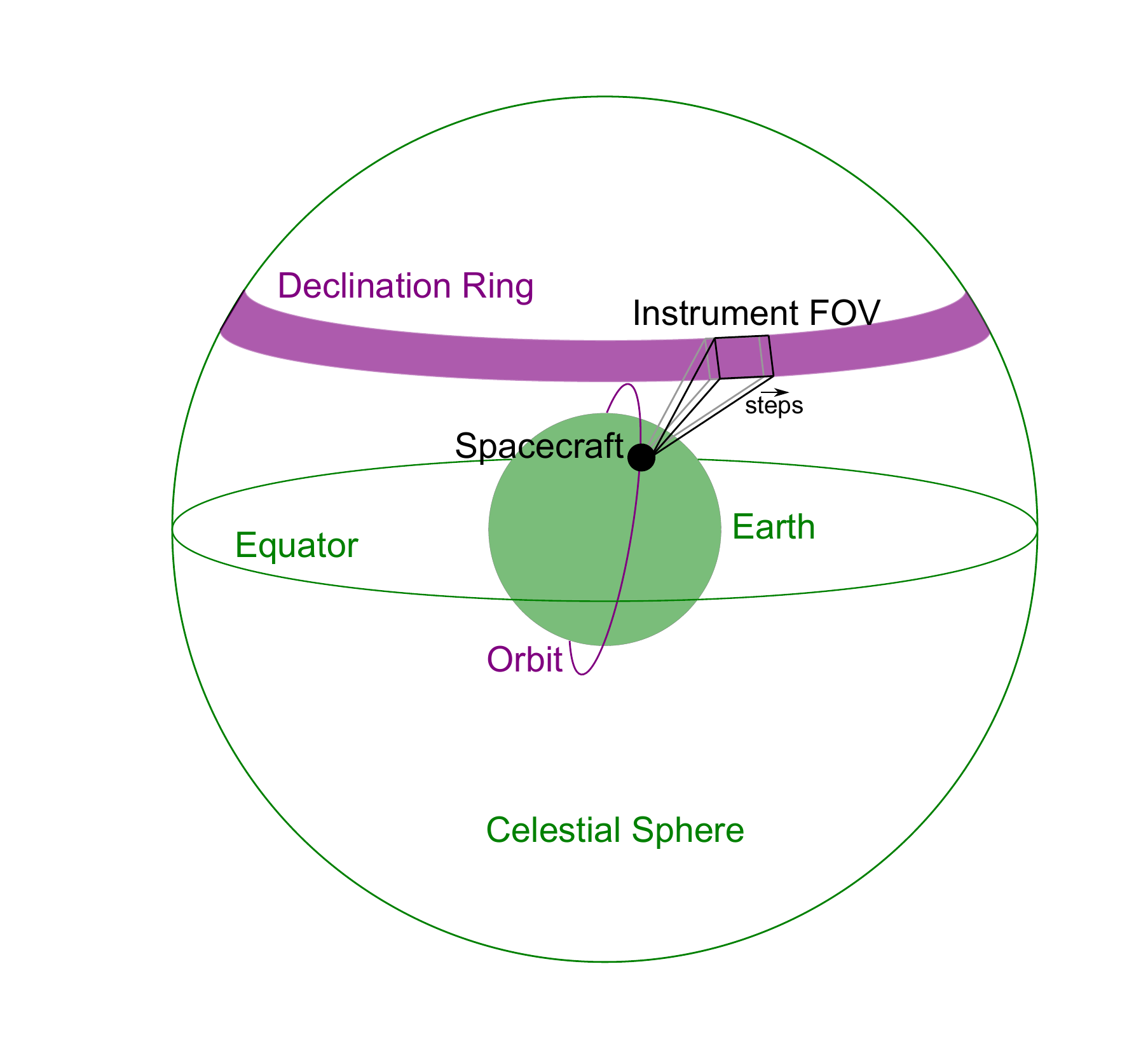}
     \end{center}
     \caption{All-sky survey strategy where the purple ring shows coverage along a ring of constant declination, where on successive pointings, the FOV is stepped along this ring. \label{fig:All_Sky_Steps_general}}
    \end{figure}

To achieve complete coverage of the all-sky survey, pointings are dedicated to certain declinations, and successive steps at that declination are taken varying the RA (where the step size is equal to the desired FOV step size).
Over time, the individual images will stack up and achieve full coverage of each declination ring, as in Fig. \ref{fig:All_Sky_Steps_general}.
The all-sky pointings are scheduled according to a heuristic-based scheduling approach, which can be applied to any time horizon.
% (a week or month).
The number of total required steps at each declination is a function of this time horizon.
%, as in Fig. \ref{fig:steps_vs_dec}.
%, gener a day or week at a time.
%
The strategy is to first identify the times that are not already scheduled by targeted pointings
%(by NCP/ SCP or Galactic Plane pointings) 
and the required slew times between pointings.
Second, the algorithm schedules each pointing by determining the maximum feasible observing time for that pointing (given a starting time), which is the minimum of the time until the next scheduled pointing and $T_{max}$ as constrained by $\beta$.
%, as in Fig. \ref{fig:Tmax_yr}.
%
Third, the declinations are identified that can be feasibly viewed for this maximum time (i.e. do not violate the Earth constraint at the start or end of the time interval).
Fourth, the feasible all-sky declination with the maximum number of remaining required steps is selected.
This prioritization enables all the all-sky declinations to be covered with the required number of steps in the given time horizon.
Fifth, the starting and ending times of this pointing are assigned, all counters are updated, and the start time of the next pointing is assigned.
This process is repeated until every block of unassigned time is scheduled with all-sky pointings.

%Prior to describing this algorithm 

The following definitions are necessary to understand the algorithm pseudo-code, 
\begin{itemize}
\item $T_s$: Vector of chronologically sorted starting times of all scheduled pointings.
\item $T_e$: Vector of chronologically sorted ending times of all scheduled pointings.
\item $d(t)$: Vector of spacecraft declination, which is a function of time, $t$.
\item $D$: Vector of required declinations for the all-sky survey.
%, which depends on the duration of time over which the scheduling is performed.  For example, $D$ for one day is plotted in Fig. \ref{fig:pts_vs_dec}
%
\item $RD$: Vector of required number of steps to cover every declination in $D$, which depends on the duration of time being scheduled.  The number of steps to be scheduled for one day is a function of declination.
%is given in Fig. \ref{fig:steps_vs_dec}.
\item $SD$: Vector of the already scheduled number of steps for every declination in $D$.
\item $PD$: Vector of chronologically sorted declination index for scheduled pointings, corresponding to the times in $T_e$ and $T_s$ with indices referring to the declinations in $D$. 
\item $n= |T_s| = |T_e|= |PD|$: Number of scheduled pointings.
\item $s$: Temporary variable indicating the number of steps in that pointing.
\item $t_{sp}$: Duration of large slew between successive pointings.
\item $t_{ss}$: Duration of small slew between successive steps.
\item $t_{st}$: Duration of small step.
\item $\tau$: Temporary time variable representing the start of the next pointing.
\item $T_{max}(\beta)$: the maximum pointing duration, which depends on $\beta$, as in Fig. \ref{fig:Tmax_yr}.
\item $\delta_{max}(\beta)$: the maximum tilt in the cross-track direction, which depends on $\beta$, as in Eq. \ref{eq:delta}.
\item $F \subset D$: Vector of feasible declinations for a given scheduling instance.
\end{itemize}

%This algorithm to schedule all remaining all-sky pointings is expressed in pseudo-code below, 
Prior to starting the all-sky scheduling algorithm, the scheduled targeted observations captured in $T_s$, $T_e$, and $PD$, and the pointing counter is  initialized to $j = n$.
The algorithm for assigning the all-sky survey pointings is as follows:
%\newpage
\begin{algorithmic}
\FOR{i=1 \TO i = n-1 } 
\STATE{ \IF{$T_s(i+1)-T_e(i) \geq t_{sp} +t_{st}$} 
\STATE{
$\tau = T_e(i) + t_{sp}$ %\label{st:1}
\WHILE{$\tau \leq T_s(i+1)-t_{sp} -t_{st}$} 
\STATE{$j \rightarrow j+1$ \\ 
$\Delta t \rightarrow \min(T_{max}(\beta), T_s(i+1)-\tau -t_{sp})$ \\
$F \rightarrow intersect( find(|d(\tau)-D|< \delta_{max}), find(|d(\tau+\Delta t)-D|< \delta_{max}))$ \\ 
$ j \rightarrow \max(RD(F)-SD(F)) $\\
$s  \rightarrow ceiling({(\Delta t+t_{ss})}/{(t_{st}+t_{ss})})$ \\
$ SD(F(j)) \rightarrow SD(F(j))+s $\\
$PD(j) \rightarrow D(j)$\\
$T_s(j) \rightarrow \tau $\\
$T_e(j) \rightarrow \tau +\Delta t$ \\
$\tau \rightarrow \tau +\Delta t + t_{sp}$  } 
\ENDWHILE}  \ENDIF
} 
\ENDFOR
\end{algorithmic}

The all-sky algorithm above will append vectors $T_s$, $T_e$, $PD$, which will no longer be ordered chronologically because the algorithm fills in time periods without scheduled pointings (however they can be easily sorted).
%
%However, this can be easily sorted to restore the chronology if desired.
%
This scheduling approach, including
%including assigning the Deep, Galactic Plane, and all-sky pointings should be iterative until the appropriate parameters are determined to achieve the scheduling goals.
%
selection of the cadence and distribution of the targeted observations,
%the Deep survey observations, 
as well as integration times for all surveys, should be iterative until the appropriate parameters are determined to achieve the scheduling goals.
In particular, they should be selected such that all required all-sky pointings are accomplished (i.e. $SD=RD$), and that the desired trade-off between scheduling efficiency and maximizing the deep survey redundancy is achieved.
%
%When the cadence and distribution of Galactic Plane and Deep survey poitings are selected correctly, the required number of steps will be scheduled, .
%
%Additional details can be added to augment and improve the above high-level schedule, for example, if there are no remaining declinations that are feasible for the entire duration $\Delta t$, we shorten the duration to find a feasible declination.
%
After the algorithm has been completed, the resulting schedule can also be improved, for example short periods of time that are not scheduled can be scheduled by increasing the number of steps (adding redundancy) or scheduling  other spacecraft operations (e.g. downloads, reaction wheel de-saturation).
% and the schedule is feasible.
%

%Describe why this results in feasible solutions and works overall.

%  \begin{figure}[!h]
%   \begin{center}
%       \subfigure[Mollweide Equal Area Projection]{
%      \includegraphics[width=.45\textwidth,angle=0]{figs/Mollweide_projection_SW}
%       \label{fig:Mollweide_projection_SW}}
%%
%       \subfigure[Maximum Pointing Time, $T_{max}$ in Orbital Plane]{
%        \includegraphics[width=.45\textwidth,angle=0]{figs/steps_vs_dec}
%        \label{fig:steps_vs_dec}}
%   \end{center}
%   \caption{Properties of all-sky Survey \label{fig:all_sky}}
%  \end{figure}

Completing the all-sky survey can be visualized as imaging a grid of the celestial sphere, with vertical bands with height equal to the instrument's FOV and within each band scanning at angular steps equal to the angular wavelength steps ($s_{AS}$).
When projected as an equal-area Mollweide map, as in Fig. \ref{fig:rep_results}, the circumference at declinations is different- maximum at the equator and minimum at the poles due to the equal-area projection.
To achieve global coverage with a redundancy of one in 6 months, a circumference of $c=360^{\circ} \cos(d)$ must be covered for each declination, $d$.
Thus, fewer steps are required near the poles relative to near the equator.
% higher and lower declinations to cover an equal-area.
%, as shown in Fig. \ref{fig:steps_vs_dec}, note this curve is similar to the vertical curvature of the Mollweide map.
%
The total number of steps to cover this area depends on the FOV size and number of required steps across the detector.
%is $s_{AS} = c/a_{AS}$, where $a_{AS}$ is the angular size of a single wavelength step.
%, as defined in Table \ref{tab:bands_overview}.
%

%This does not enforce any addition
%%
%Something about rounding to the nearest number of steps and depending on the duration being scheduled.
%%
%Discuss implications on observing scenario and total number of steps per orbit, but how this isn't really a concern b/c already satisfied with the required number of steps and pointings.
% including discretization of latitudes for all-sky survey
% 
%Explain the differential precession experieinced at different latitudes why there are more pointings centered at equator (maybe plot showing why).
%%
%Discuss how there's a list of pointings to be completed based on the declination.
%%
%Discuss high and low latitude pointings, where Deep (NCP/SCP) pointings are not also possible.
%%
%
%Describe minimum number of steps per orbit to achieve redundancy requirements.
%
%Discuss overlap from GP pointings.

\subsection{Special Case: Galactic Plane Survey}
\label{sec:gal_case}

To demonstrate how a targeted observation can be accommodated in the observing strategy and due to the scientific interest in the Galactic plane, we address the special case where in addition to the all-sky survey, the mission must include coverage of the Galactic plane
%is covered in the all-sky survey, however a different strategy may be required in the case that 
with higher spatial resolution or redundancy than the rest of the survey.
% is considered.
%
%This can be accommodated with some modifications to the overall strategy.
% to achieve the desired coverage over one degree above and below the Galactic Plane with the desired redundancy.
%
In a LEO, the RA where the spacecraft crosses the Galactic plane varies throughout the year, so the orbit declination where the orbit crosses the Galactic plane will be dynamic.
%see Fig. \ref{fig:gal_long_GPcross}, 
The spacecraft has two opportunities to image every part of the Galactic plane every year due to the orbit precession, which provides some flexibility on when Galactic plane pointings are scheduled to satisfy its redundancy requirements.

\begin{figure}[!h]
 \begin{center}
     \includegraphics[width=.4\textwidth,angle=0]{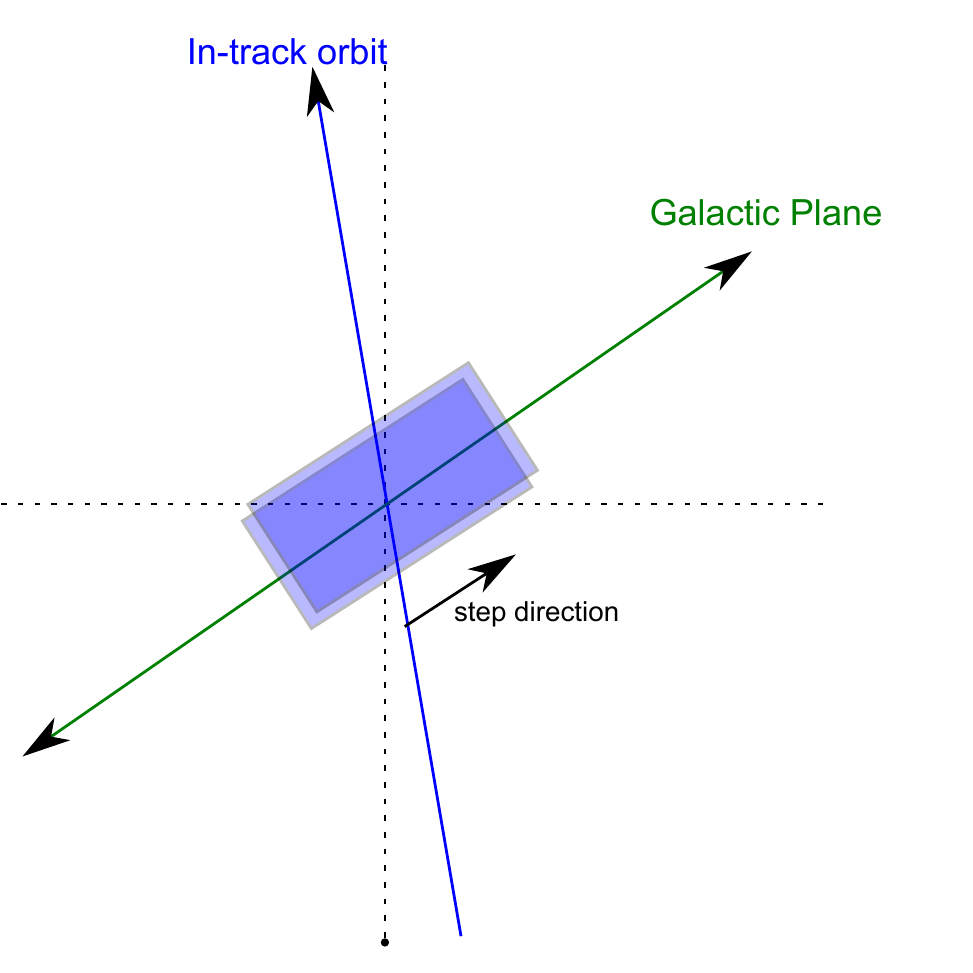}
 \end{center}
 \caption{Steps along Galactic plane to achieve complete coverage \label{fig:GP_images_steps_v1}}
\end{figure}

The Galactic plane runs at an angle relative to the orthogonal lines of RA and declination.
%, see Fig. \ref{fig:al_sky_Mollweide}.
%
%To achieve full wavelength coverage, 
To efficiently cover this area, the proposed strategy is to rotate the spacecraft such that the long end of the FOV is parallel to the Galactic plane and the FOV is centered over the Galactic plane.
Thus, successive steps along the Galactic plane with step size to achieve the specific survey coverage goals, as in Fig. \ref{fig:GP_images_steps_v1}.
The number of steps required to cover the Galactic plane depends on the FOV and redundancy requirements.
Note the ability to rotate the spacecraft may be constrained by solar panel or star tracker angles, or thermal limitations.

%In reality, there is a Galactic Plane pointing some fraction of orbits ranging from 0.4-0.6, depending on the number of steps per pointing (constrained by $T_{max}$, depending on $\beta$).
%
%The Galactic plane pointing declination varies throughout the year as in Fig. \ref{fig:gal_long_GPcross}.
%

%\newpage
\section{Application to the SPHEREx Mission}
\label{sec:appl_SPHEREx}

This section applies the scheduling strategy and algorithm to the SPHEREx
% (Spectro-PHotometer for the Extragalactic structure, Reionization and ices Explorer)
   mission,
   % a small explorer (SMEX) mission 
   proposed by California Institute of Technology (Caltech) and Jet Propulsion Laboratory (JPL).
   % and Ball Aerospace and Technologies, Corp. (BATC).
  %
  The SPHEREx is an astrophysics mission performing an all-sky spectroscopic survey, studying inflationary cosmology, the history of galaxy formation, and Galactic ices.
  The SPHEREx spacecraft will be in launched into a 500 km altitude sun-synchronous (inclination=$97.4^{\circ}$)  nearly terminator orbit (18 hr orbit) with a period of 94.6 minutes, selected to minimize thermal concerns and maximize power collection with the ability to view the entire celestial sphere.
This two year mission has requirements to cover the entire celestial sphere once every six months, the Galactic plane once every six months, and maximize coverage of the North Celestial Pole (NCP) and South Celestial Pole (SCP) regions.
  % (TBC).
  %TBD- provide more info on the science or reference to another paper.
  %

In a nearly-terminator Sun synchronous orbit, the $\beta$ angle  varies between $60^{\circ}-90^{\circ}$ throughout the year, see Fig. \ref{fig:beta_yr}.
  % (defined in Section \ref{sec:prob_desc}.\ref{sec:defs})
The SPHEREx pointing constraints are: $\alpha \leq 35^{\circ}$, $\zeta \leq 35^{\circ}$, $\Omega=90^{\circ}$ thus $T_{max}$ varies from 9 to 19 minutes throughout the year, see Fig. \ref{fig:Tmax_yr}.
For the SPHEREx spacecraft, $\phi$ is a fixed angle that must be determined before the spacecraft is developed (and impacts other subsystems such as thermal and attitude determination and control), while $\theta$ is a free decision variable in the scheduling problem that can be dynamic over time.

%    \begin{figure}[!h]
%     \begin{center}
%         \includegraphics[width=.6\textwidth,angle=0]{figs/al_sky_Mollweide}
%     \end{center}
%     \caption{Three SPHEREx surveys projected on celestial Mollweide sky map.  The blue polar regions shows the Deep survey regions, the red line indicates the galactic plane region, and the white zone indicates the all-sky survey regions. \label{fig:al_sky_Mollweide}}
%    \end{figure}
  
\subsection{Survey Overview}

\begin{figure}[!h]
 \begin{center}
     \includegraphics[width=.6\textwidth,angle=0]{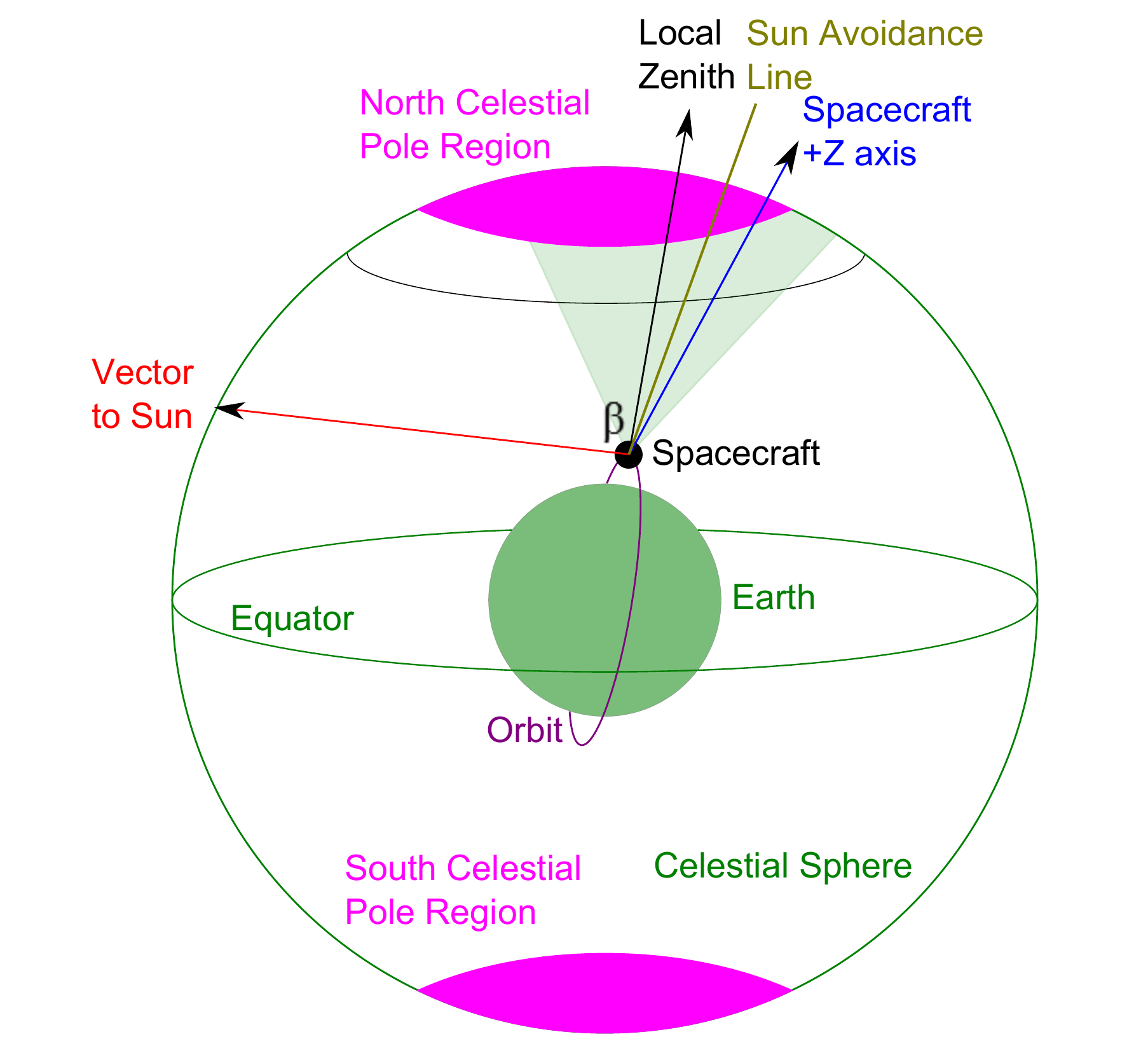}
 \end{center}
 \caption{Constraints on spacecraft orientation for the SPHEREx mission. \label{fig:3D_constraints_spherex}}
\end{figure}

The three surveys focus on different areas of the celestial sky
%as shown in Fig. \ref{fig:3D_constraints_spherex}, 
and have different integration and redundancy requirements.
The deep survey consists of surveying 100$^{o}$ both around the NCP and SCP, respectively (which can be expressed as all declinations $\leq 83.5^{\circ}$ and $\geq 83.5^{\circ}$), as in Fig. \ref{fig:3D_constraints_spherex}.
The NCP and SCP are selected as the regions where deep surveys are done because the poles are the natural rotation axis for the SPHEREx orbit and the instrument can access these regions throughout the year.
The Galactic plane survey requires coverage of approximately one degree above and below the Galactic plane.  
%
%In reality about half the in the field of view (FOV)
%redun
The all-sky survey consists of the remaining celestial sphere and the full data set will use data from the Deep and Galactic surveys to achieve full celestial sphere coverage.

%

%\begin{figure}[!h]
% \begin{center}
%     \includegraphics[width=.5\textwidth,angle=0]{figs/sky_map}
% \end{center}
% \caption{Three SPHEREx Surveys on celestial sky map.  \label{fig:sky_map}}
%\end{figure}

\begin{figure}[!h]
 \begin{center}
     \includegraphics[width=.6\textwidth,angle=0]{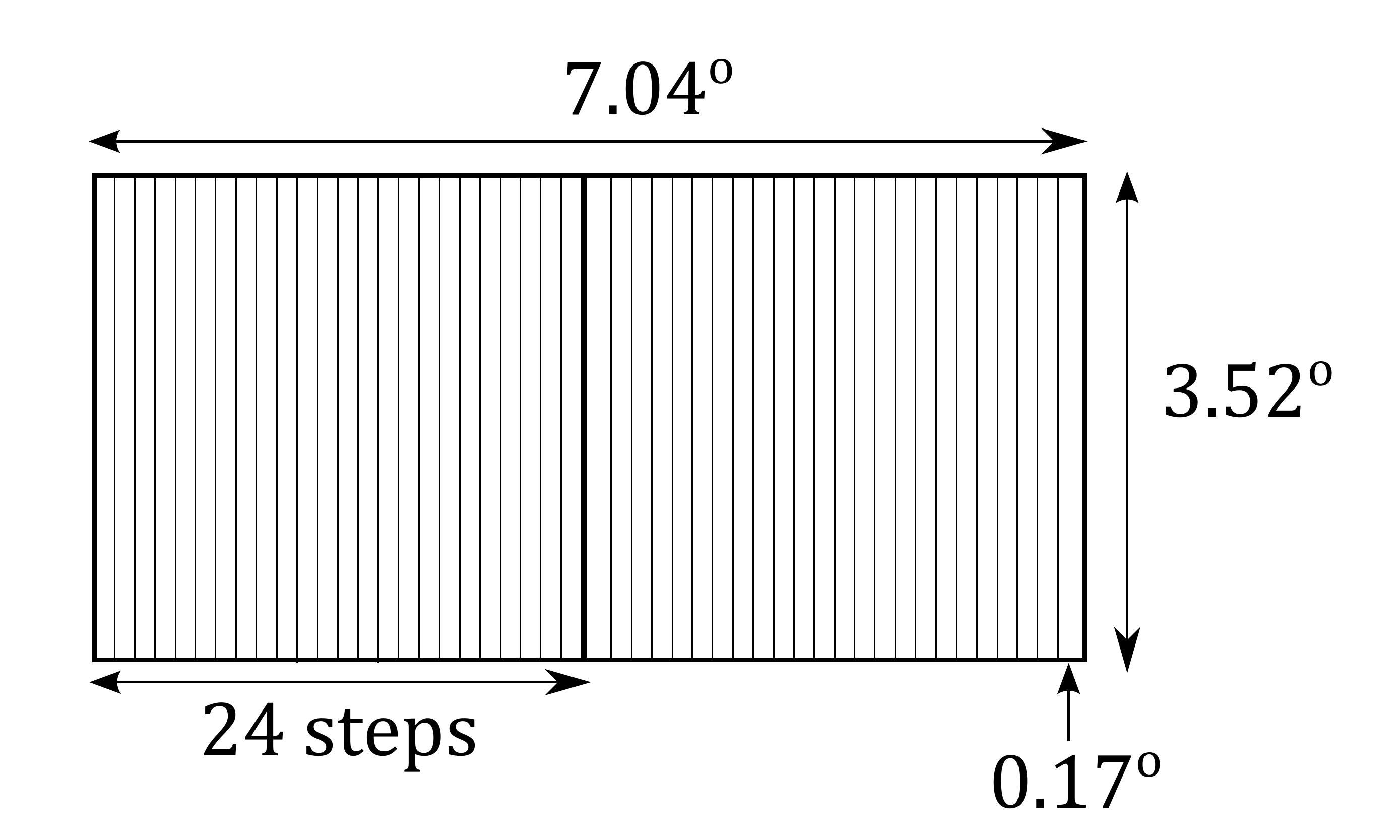}
 \end{center}
 \caption{Field of view (FOV) with 24 steps across each detector (as used for the all-sky and Galactic plane surveys\label{fig:FOV_steps} }
\end{figure}

The SPHEREx instrument has a field of view (FOV) that is $7.04^{o}$ by $3.52^{o}$, which consists of two size-by-side detectors that are each $3.52^{o}$ square, see Fig. \ref{fig:FOV_steps}.
%  
%
%An image of the orbit and the Zenith-pointing FOV is shown in Fig. \ref{fig:3D_constraints}.
%
There are also two detectors stacked behind the two detectors in the image for a total of four detectors.
The detectors are Linear Variable Filters (LVFs), which are wedge filters, where the thickness and thus wavelength varies continuously along one dimension, the scan direction \cite{Rosenberg}.

%%in Table \ref{tab:survey_overview} 
%  \begin{figure}[!h]
%   \begin{center}
%       \subfigure[Field of view (FOV) with 21 steps across each detector (as used for the all-sky and galactic plane surveys ]{
%      \includegraphics[width=.6\textwidth,angle=0]{figs/FOV_steps}
%      %sky_map}
%       \label{fig:FOV_steps}}
%%
%\hspace*{4em}
%%
%       \subfigure[SPHEREx FOV when near NCP]{
%        \includegraphics[width=.25\textwidth,angle=0]{figs/FOV_view}
%        \label{fig:FOV_view}}
%   \end{center}
%   \caption{SPHEREx Map and FOV \label{fig:sky_FOV}}
%  \end{figure}

The surveys have different wavelength and spatial resolution requirements, summarized in Table \ref{tab:survey_overview}.
% with the wavelength bands defined in Table \ref{tab:bands_overview}.
%
Every survey needs to have complete wavelength coverage, meaning every wavelength (long rectangles in Fig. \ref{fig:FOV_steps}) FOV covers every area in the survey.
Wavelength coverage is achieved by stepping the detector by the step required to achieve the required spectral resolution.
%, see Table \ref{tab:bands_overview}.
%
In this mission application, redundancy is defined as the number of times a given area of sky is covered by every wavelength band applicable for that survey.
The Galactic and all-sky surveys require a redundancy of once per six months, meaning each half of the orbit (ascending and descending) covers the entire celestial sphere in view as it precesses.
The scheduling objective is to maximize the redundancy of the Deep surveys, i.e. uniformly sample the NCP and SCP and surrounding areas the maximum number of times, subject to the requirements of all surveys.

\begin{table}[!h]
   	\caption{SPHEREx Survey Overview \label{tab:survey_overview}}
   	\centering
       	\begin{tabular}[!h]{|l|c|c|c|}
   		\hline \hline
Survey	&	Deep (NCP/ SCP) 	&	Galactic Plane  & All-Sky	\\ \hline \hline
Parts of celestial sphere & NCP/ SCP (100$^{\circ}$) &  1$^{\circ}$ of Galactic plane & All remaining sky  \\ \hline
%
%Bands	&	1-3	&	4 & 1-3	\\ \hline
Spectral Resolution & 40 & 40 & 40 \\ \hline 
Steps Across the Detector & 21 & 21 & 21  \\ \hline
Integration Time (per step) & 185 secs & 95.7 secs & 95.7 secs	\\ \hline
Redundancy Requirements & Maximized & once per 6 months & once per 6 months	\\ \hline
    \end{tabular}
\end{table}

\begin{table}[h]
   	\caption{SPHEREx Schedule Parameters\label{tab:params}}
   	\centering
       	\begin{tabular}[!h]{|l|c|c|}
   		\hline \hline
Parameter	&	Values	&	Units	\\ \hline \hline
Orbits per day	&	15.2	&	orbits/day	\\ \hline
Orbits per year	&	5552	&	orbits	\\ \hline
Orbit Period	&	94.7	&	mins	\\ \hline
Orbit Precession	&	1	&	deg/day	\\ \hline
FOV half width	&	3.52	&	deg	\\ \hline
all-sky/ Galactic Integration Time	&	95.7	&	sec	\\ \hline
Deep (NCP/SCP) Integration Time	&	185	&	sec	\\ \hline
Small Slew Duration (between steps)		& 10	&	sec	\\ \hline
Large Slew Duration (between pointings)	&	90	&	sec	\\ \hline
Telecom Time & 1.6 & min/orbit \\ \hline
    \end{tabular}
\end{table}

The SPHEREx scheduling parameters are in Table \ref{tab:params}, which are a function of the orbit altitude and inclination, FOV, and preliminary science, scheduling, and attitude control calculations.
Telecommunication operations requires an average of 1.6 minutes per orbit, which is scheduled for about seven minutes every three to four orbits.
This telecommunication time is accounted for in the overall scheduling.

%Steps (bands) across FOV	&	20.5	&	steps	\\ \hline
%Angular coverage per step	&	0.171707317	&	deg	\\ \hline
%	&	10.30243902	&	arcmin	\\ \hline
%Steps (bands) across FOV	&	98.1	&	steps	\\ \hline
%Angular coverage per step	&	0.035881753	&	deg	\\ \hline
%	&	2.152905199	&	arcmin	\\ \hline

\subsection{Spacecraft Configuration}

The instrument is at a fixed cant angle, $\phi$, relative to the spacecraft (i.e. it cannot move dynamically throughout the orbit or year), which is a design variable that interacts with the observing scenario.
%, is a free design variable.
%
In order for the instrument to have the NCP and SCP in view throughout the entire year which is necessary to maximize the Deep survey redundancy, and considering the tilt angle for solar avoidance ($\theta$), $\phi=21^{\circ}$ was selected. 
There is one limiting case throughout the year where either the NCP or SCP is on the edge of the FOV (i.e. a single wavelength range can image the NCP or SCP).
A Sun-synchronous orbit with a longitude of descending node is 18 hours is selected such that the limiting case is during the winter solstice for the SCP ($\beta=60^{\circ}$), because the NCP is higher priority.
In the worst case for the NCP, the FOV is 2.4$^{\circ}$ from the NCP center, which occurs during the Summer Solstice ($\beta=74^{\circ}$). 
%
%TBD- add figures and more details on this decision and driving cases.
%This allows the spacecraft 
%Describe how spacecraft and instrument are fixed relative to orbit and themselves such that we can always observe the NCP/SCP and the optimal cant angle of the instrument.
%
%The Sun-avoidance constraint does not directly constrain th
%
%Discuss why an 18 hr orbit was selected instead of a 6 hr orbit.

%\subsection{Schedule Parameters}

%Describe how three key constraints- Sun, Earth, and Moon, really drive scenario.
%%
%Describe general approach of keeping spacecraft  pointing roughly in the zenith direction and in the orbital plane and how this has to satisfy both Sun and Earth constraints.
%

  \begin{figure}[!h]
   \begin{center}
       \subfigure[Solar $\beta$ Angle]{
      \includegraphics[width=.45\textwidth,angle=0]{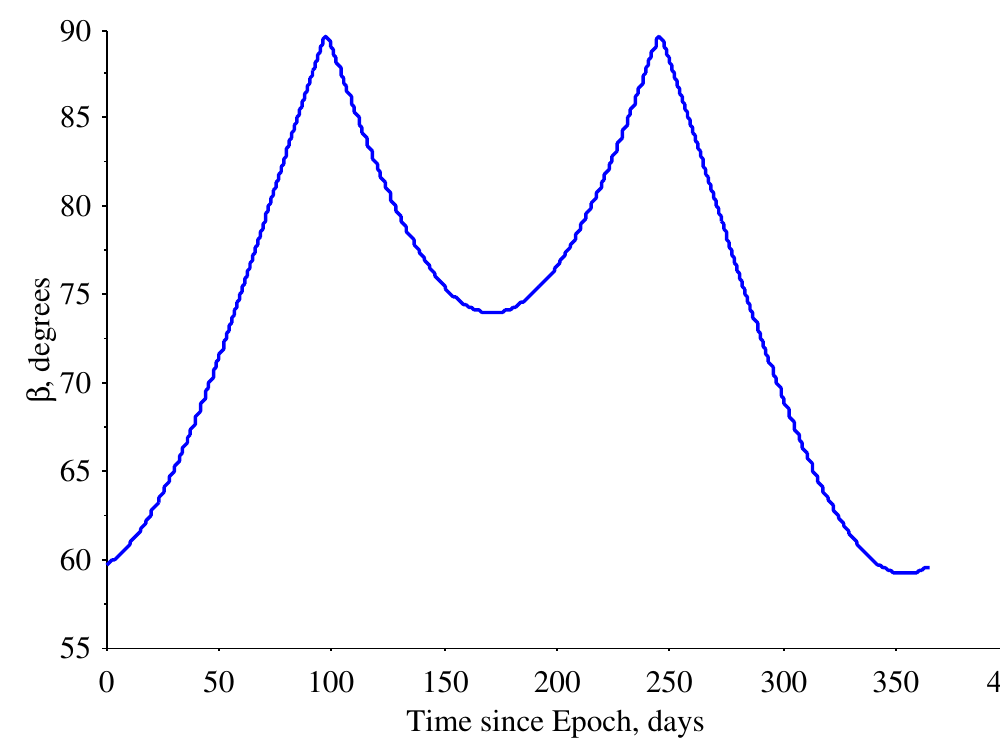}
       \label{fig:beta_yr}}
       \subfigure[Maximum Pointing Time, $T_{max}$ in Orbital Plane]{
        \includegraphics[width=.45\textwidth,angle=0]{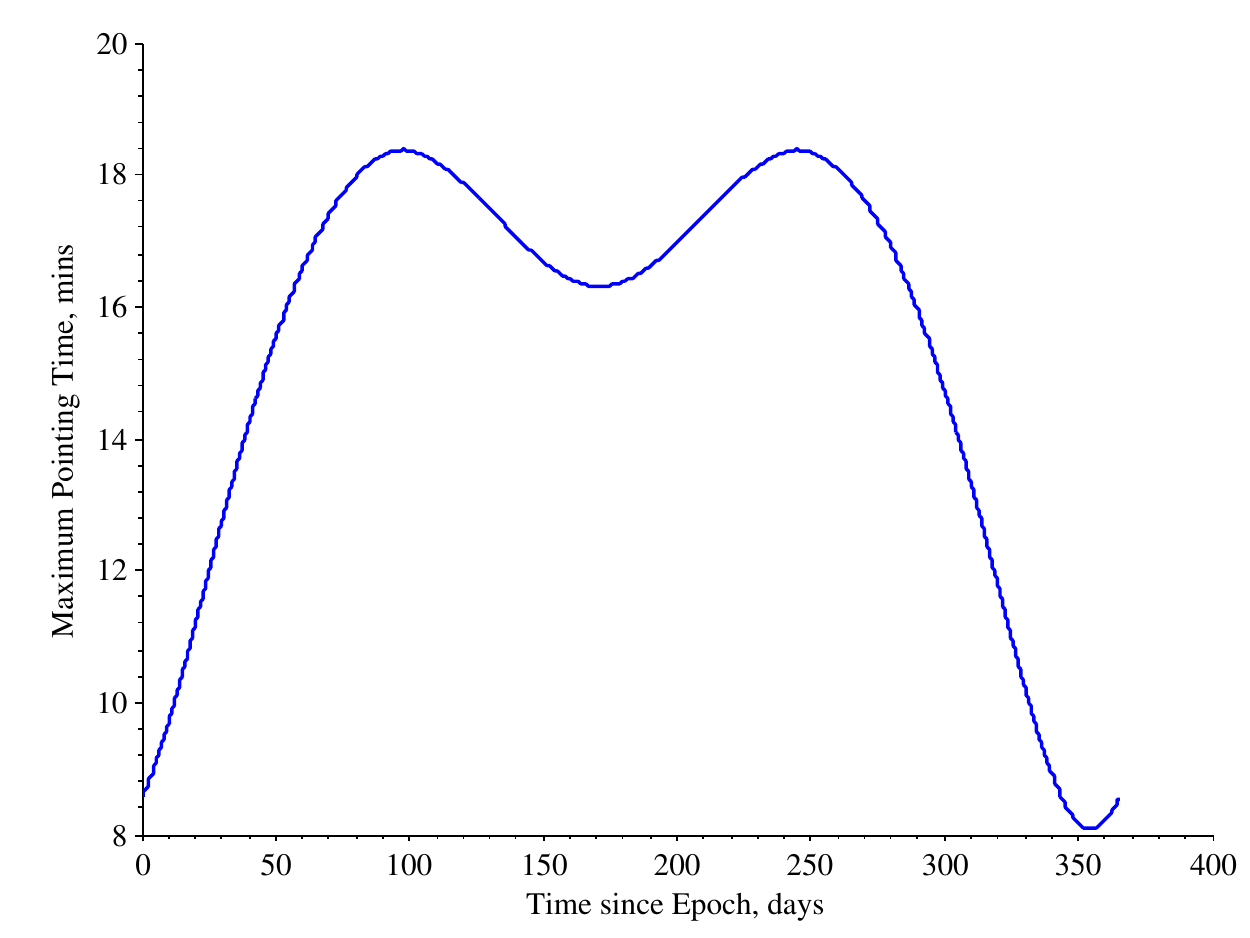}
        \label{fig:Tmax_yr}}
   \end{center}
   \caption{The maximum pointing time is a function of the solar $\beta$ angle as in Eq. \ref{eq:Tmax}.  \label{fig:beta_Tmaxl}}
  \end{figure}

For the $\beta=90^{\circ}$ case, generally the angular difference between successive slews is about $60^{\circ}$ ($360^{\circ}/6$ pointings).
%, due to the fact that we do nominally 6 pointings per orbit.
%
Differences in successive pointing angles of up to approximately $80^{\circ}$ can be tolerated (depending on the declinations), which exceeds the $75^{\circ}$ that may be expected due to tilting from the local Zenith by $35^{\circ}$ in each direction because of the motion of the spacecraft during the slew (about 90 seconds).
% (when $\beta=90^o$).

%Describe how constraints lead to the strategy: general division and strategy to accomplish three surveys, including discretization of latitudes for all-sky survey, rotation of spacecraft/instrument to scan along Galactic Plane, and general approach for Deep (NCP/SCP) survey.
%%
%

%
%Discuss why we might sometimes have 5 pointings....

%\vspace{-5mm}
%\begin{table}[h]
%	\caption{SPHEREx Survey Bands  \label{tab:bands_overview}}
%	\centering
%    	\begin{tabular}[!h]{|l|c|c|c|c|}
%		\hline \hline
%Bands & Wavelength ($\mu m$) & Spectral Resolution & Steps Across Detector & Angular Step Size	\\ \hline \hline
%Units & $\lambda $ & r & $s_{AS}$ & $a_{AS}=3.52^{\circ}/s_{AS}$ \\ \hline 
%1 & 0.75-1.25 & 40 & 21	& 0.17$^{\circ}$ \\ \hline 
%%
%2 & 1.25-2.09 & 40 & 21	& 0.17$^{\circ}$\\ \hline 
%%
%3 & 2.09-3.5 & 40 & 21 & 0.17$^{\circ}$	\\ \hline 
%%
%4 & 3.5-5.0 & 150 & 98 & 0.036$^{\circ}$	\\ \hline 
%    \end{tabular}
%\end{table}

\subsection{Strategy Overview}

The SPHEREx mission is considered an ``engineered'' survey because the three survey goals are a natural synergy enabled by its polar orbit.
% Low Earth Orbit (LEO).
%
Removing one of the surveys would not directly impact the other surveys, and in fact, as a result, the spacecraft may be idle for portions of the orbit.
On every orbit the instrument can access the NCP and SCP (enabled by the selection of appropriate cant angle $\phi$), the Galactic plane, and a band of the celestial sphere at constant right ascension (RA) for all-sky survey coverage.
In general the surveys do not directly conflict because they target different areas of the celestial sphere, however there are trade-offs between the various surveys. 
%, but the goal is to satisfy the requirements of all three surveys in an efficient way.

    \begin{figure}[!h]
     \begin{center}
         \includegraphics[width=.7\textwidth,angle=0]{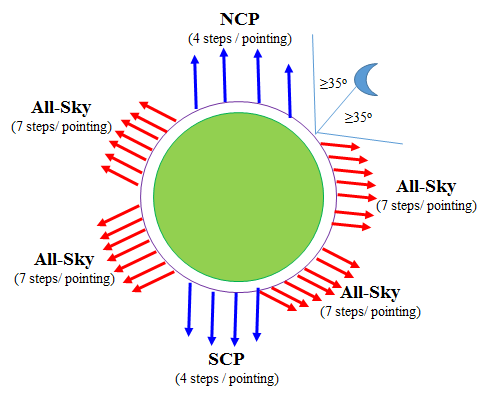}
     \end{center}
     \caption{General observing scenario for $\beta= 90^{\circ}$ case where there are 6 pointings per orbit. \label{fig:nom_pointings}}
    \end{figure}
    
The general strategy to maximize science time is to minimize the number of pointings because the large slew durations are longer than the small slew durations, see Table \ref{tab:params}.
However; the pointing durations must never exceed $T_{max}$, see Fig. \ref{fig:Tmax_yr}, so when $\beta=60^{\circ}$ at least 8 pointings are required, and when $\beta=90^{\circ}$, at least 6 pointings are required.
Fig. \ref{fig:nom_pointings} shows a representative distribution of pointings throughout an orbit for the $\beta=90^{\circ}$, where there is one NCP, one SCP, and 4 all-sky pointings (which also cover the Galactic plane) per orbit.
In this example, the number of arrows represents the number of steps per orbit, and an example is given to satisfy the Moon avoidance constraint.
% (discussed further in Section \ref{sec:obs_str}.\ref{sec:moon_avoid}).
%
Every pointing consists of several small steps, where the step direction depends on the survey type.
%generally taken in the RA direction.
%
The distribution of number and location of pointings and steps will vary depending on $\beta$, location of Galactic plane, and specific orbit.
%

%        \begin{figure}[!h]
%         \begin{center}
%             \includegraphics[width=.6\textwidth,angle=0]{figs/steps_vs_dec}
%         \end{center}
%         \caption{Required all-sky survey steps per day as a function of declination. \label{fig:steps_vs_dec}}
%        \end{figure}

To achieve full wavelength coverage of the all-sky survey, successive pointings will pick up where the last pointing left off such that the individual images stack and achieve full wavelength coverage, as in Fig. \ref{fig:All_Sky_Steps_general}.
A total of 367 All Sky steps must be accomplished per day, or approximately 24 steps per orbit to account for covering both the ascending and descending sides of the orbit.
The constraint of achieving global coverage does not introduce any more constraints on the scheduling problem.
This is because the constraints are satisfied in both extreme cases,
 when $\beta=60^{\circ}$ (with at least 8 pointings per orbit and 4 steps/pointing)
 and when $\beta=90^{\circ}$ (with at least 6 pointings per orbit and 6 steps/pointing).
However; if the redundancy requirement increases or the number of steps/ pointing changes dramatically, this may introduce new constraints.
The Galactic and all-sky surveys have the same wavelength and redundancy requirements, as in Table \ref{tab:survey_overview}, thus the Galactic science is accomplished as part of the all-sky survey.
In the case that the Galactic survey requirements differ, an alternative approach may be required to satisfy its requirements, as discussed in Section \ref{sec:obs_str}.\ref{sec:gal_case}.

To maximize the Deep Survey observations, there is an NCP and SCP pointing on a large number of orbits.
%
%The cadence for the these observations is selected to achieve the required total number of steps per time that satisfies coverage requirements and is feasible with the rest of the survey.
%
Deep survey observations are not feasible on every orbit because of the need to cover high-declination all-sky areas.
%
%per year for these two surveys, 
For example see the fractions in Table \ref{tab:schedule_beta90} for the $\beta=90^{\circ}$ case.   
Efficient coverage of the polar caps is achieved by sliding the FOV along lines of constant RA for a given number of days, and repeating the pattern at the next RA once the orbit has precessed.
% by this amount.
%
To achieve uniform coverage over the deep region, the number of steps at each declination will not be exactly even.
There will be incomplete wavelength overhang on to the all-sky survey that will augment its coverage but will not be part of the Deep survey.
%
%Describe how will aim at certain declinations and general strategy.
Note the band of constant RA is not the orbital plane because the instrument is offset from the local zenith by the combination of the static $\phi$ and dynamic $\theta$.
%(the orbit track projected onto the celestial sphere).
%

%\subsection{Idealized Schedule}

Table \ref{tab:schedule_beta90} provides an idealized summary of the breakdown for the three surveys, pointings, and steps the representative $\beta=90^{\circ}$ scenario, which is shown in Fig. \ref{fig:pointings_beta90}.
This case assumes perfect scheduling (i.e. every second is scheduled perfectly) and overall average number of pointings and steps (which is why there are fractional values).
Overall, $84.7 \%$ of the time is dedicated to science observations, with the largest fraction of other time dedicated to large slews (five 90 second slews).

\begin{table}[!h]
\caption{Representative ideal schedule overview for $\beta=90^{\circ}$ with the scenario parameter in Table \ref{tab:params}. \label{tab:schedule_beta90}}
\centering
   	\begin{tabular}[!h]{|l|c|c|c|c|c|}
\hline 
Parameter	&	Deep (NCP/SCP) &	All-Sky	&	Total	&	Units	\\
	&	 Survey	&	Survey &	&	\\ \hline \hline
Pointings per orbit	&	2	&	4	&		6	&	pointings	\\ 
(average)	&		&		&			&		\\ \hline
Pointings per year	&	11104	&	22207		&	33311	&	pointings	\\ 
 (average)	&		&		&			&		\\ \hline
Steps per pointing	&	4	&	7		&		&	steps	\\ 
 (average)	&		&		&			&		\\ \hline
Steps per orbit 	&	8	&	28		&	36	&	steps	\\ 
(images per orbit)	&		&			&		&		\\ \hline
Steps per year	&	44414	&	15545	&		199865	&	steps	\\ \hline
Integration time per step	&	185	&	99	&			&	secs	\\ \hline
Science time	&	1660	&	3156	&	4816	&		secs	\\ \hline
Percentage of science time	&	34.5	&	65.5	&		100	&	$\%$	\\ \hline
Total time 	&	28.7	&	56.6		&	85.3	&	mins	\\ 
(including small slews) &		&		&			&		\\ \hline
Large Slews	&		&		&		7.5	&	mins	\\ \hline
Telecom 	&		&		&			1.6	&	 mins \\ \hline
Total scheduled time 	&		&				&	93.6	&	mins	\\ 
(including small slews) &		&		&			&		\\ \hline
%Extra Non-scheduled Time	&		&		&			1.1	&	mins	\\ \hline
Science efficiency	&		&		&			84.7	&	$\%$	\\ \hline
\end{tabular}
\end{table}

\subsection{Scheduling Algorithm Implementation}
\label{sec:implement_scheduling}

\begin{figure}[!ht]
\begin{center}
   \subfigure[$\beta=90^{\circ}$ with deep (NCP/SCP) pointings, skipping every fourth orbit with 4 steps/ pointing.  There are usually 6 pointings per orbit due to $T_{max}=19$ mins.]{
  \includegraphics[width=.75\textwidth,angle=0]{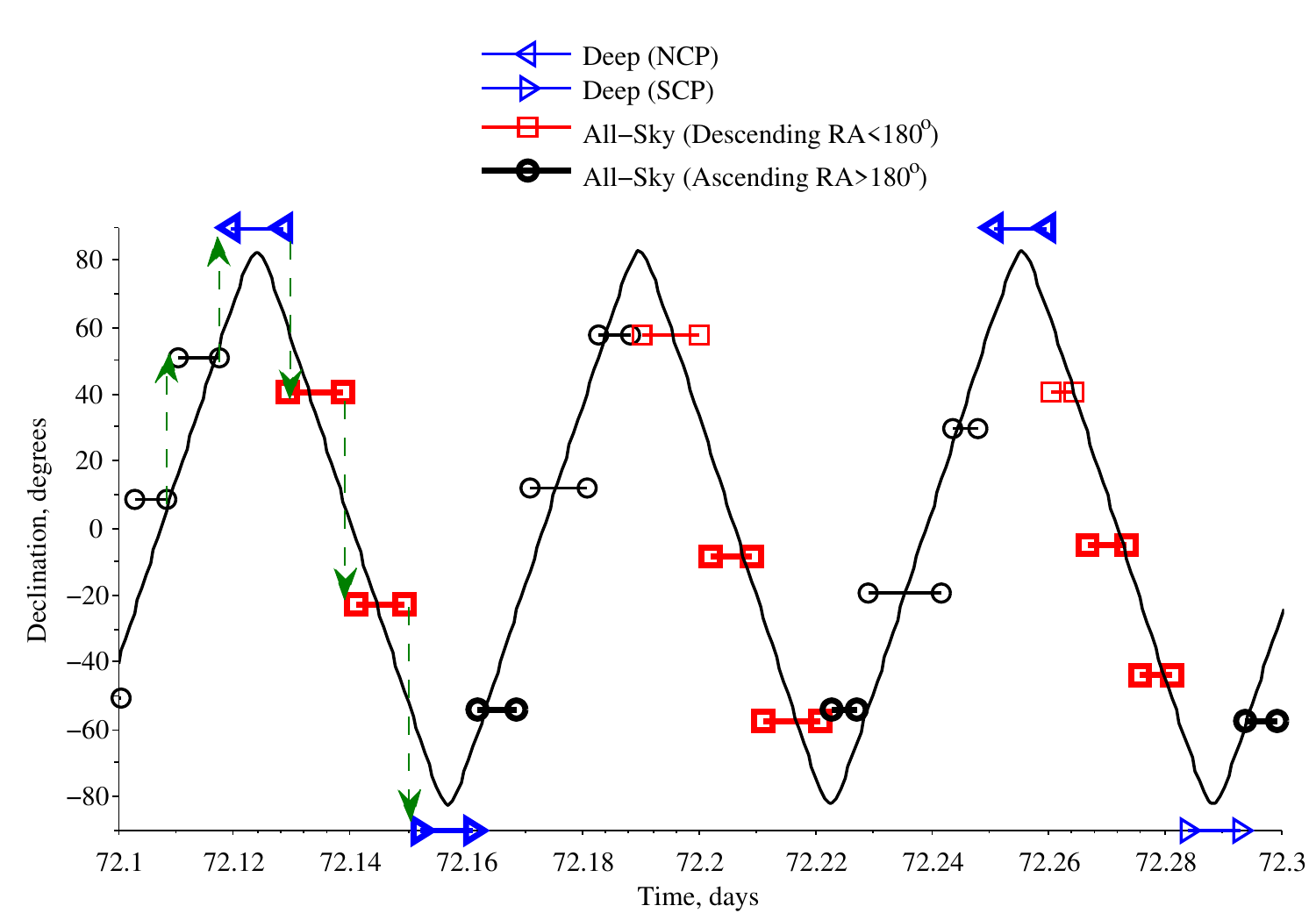}
   \label{fig:pointings_beta90_zoom}}
   \subfigure[$\beta=60^{\circ}$ with Ddeep (NCP/SCP) pointings every second orbit, with 3 steps/ pointing.  There are usually 8 pointings per orbit due to $T_{max}=9$ mins.]{
  \includegraphics[width=.75\textwidth,angle=0]{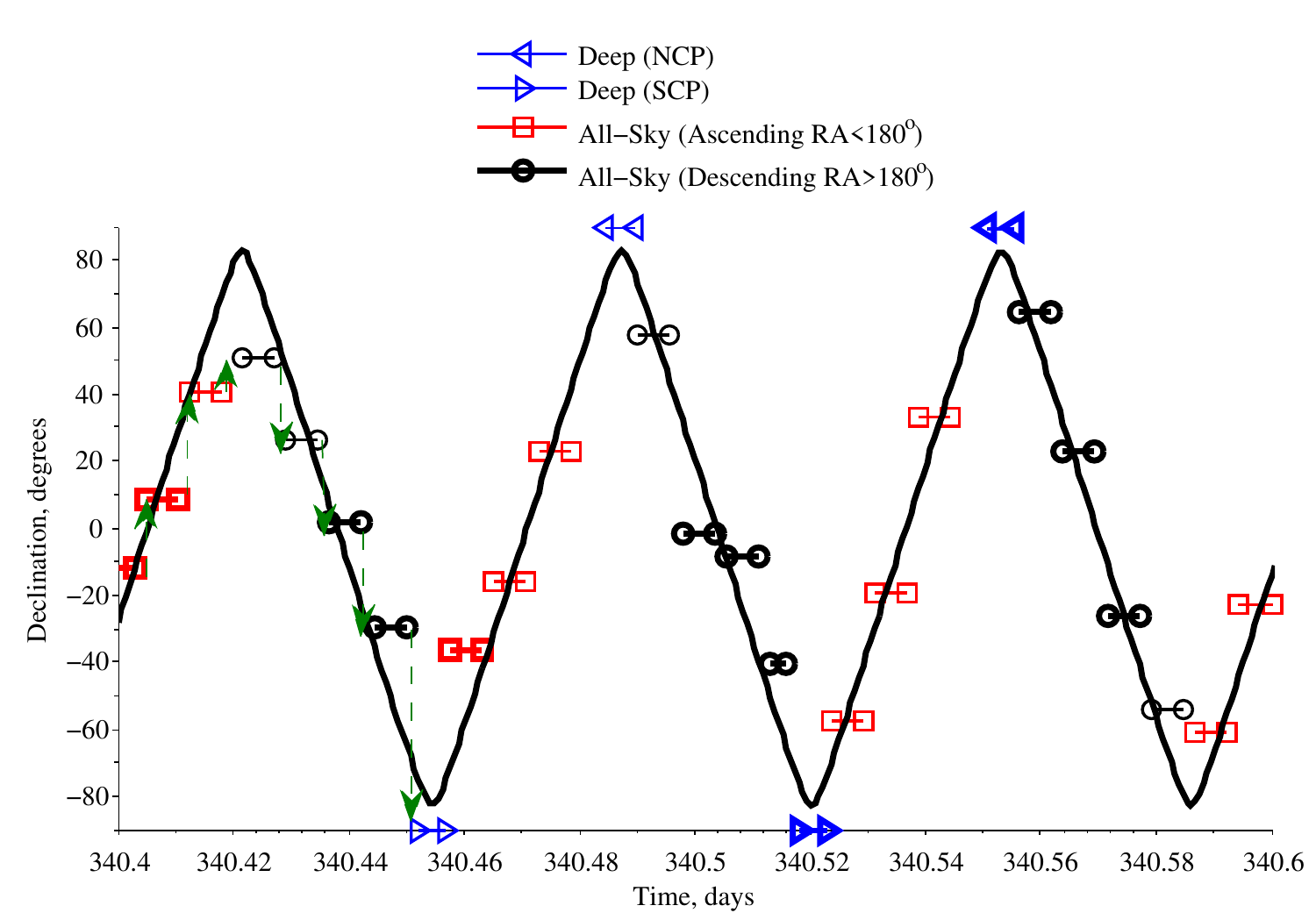}
   \label{fig:pointings_beta60_zoom}}
\end{center}
\caption{Representative schedules for three orbits (detailed view of the schedules in Fig. \ref{fig:rep_schedules}). 
  The black solid lines denote the orbit track on the celestial sphere, which establishes the available declinations as a function of time.
  The  green dotted lines with arrows denote the large slews between successive pointings.
  The symbols denote the survey pointings, which are chosen to satisfy the constraints, where each pointing is comprised of 4-9 steps where the step duration depends on the survey type.
  There are large slews (90 sec) between successive pointings are shown in green dotted lines and small slews (10 sec) between successive steps (in a single pointing) not shown here.
  \label{fig:rep_schedules_zoom}}
\end{figure}

\begin{figure}[!ht]
\begin{center}
   \subfigure[$\beta=90^{\circ}$ with deep (NCP/SCP) pointings, skipping every fourth orbit with 4 steps/ pointing.  There are usually 6 pointings per orbit due to $T_{max}=19$ mins.]{
  \includegraphics[width=.75\textwidth,angle=0]{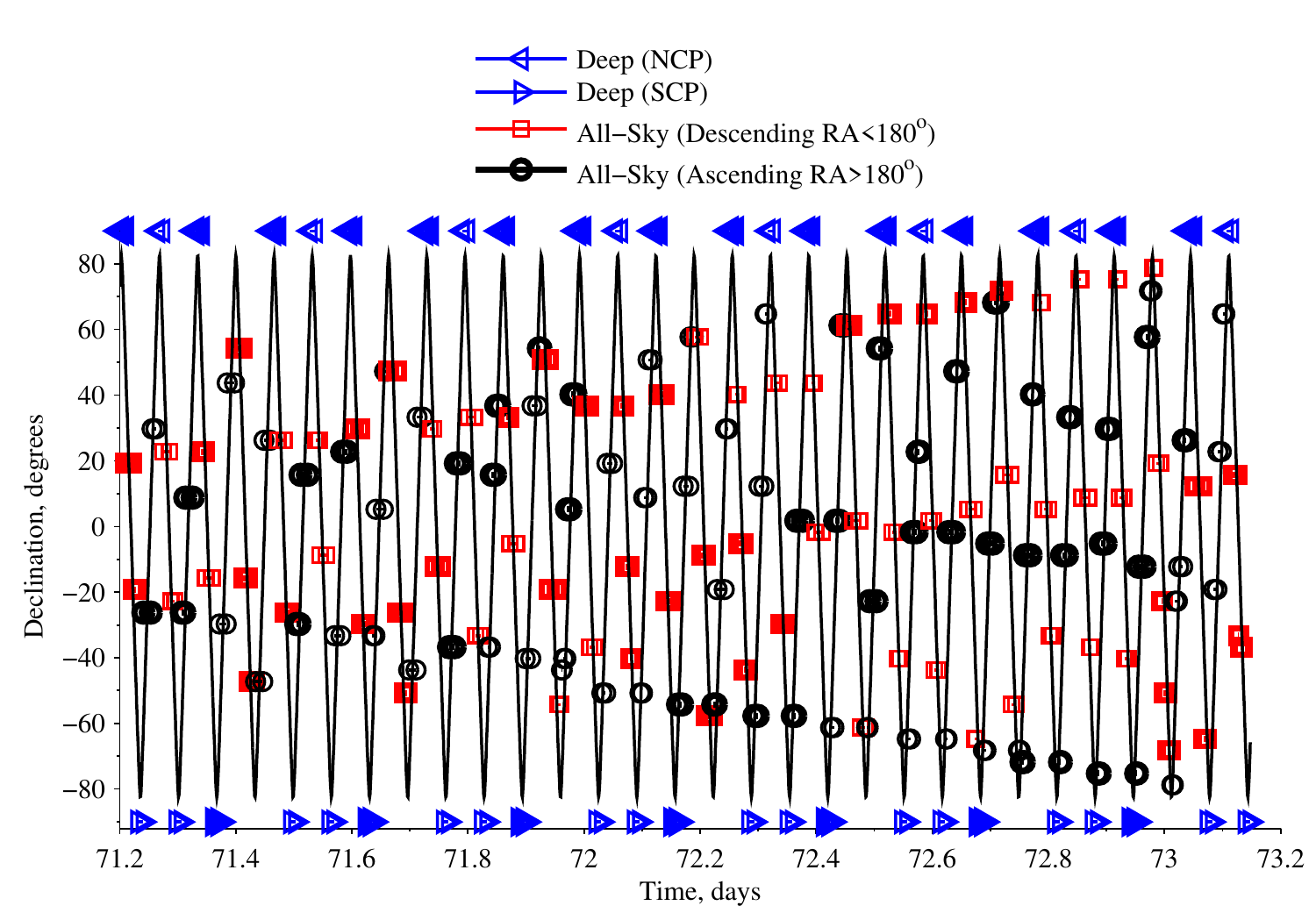}
   \label{fig:pointings_beta90}}
   \subfigure[$\beta=60^{\circ}$ with deep (NCP/SCP) pointings every second orbit, with 3 steps/ pointing.  There are usually 8 pointings per orbit due to $T_{max}=9$ mins.]{
  \includegraphics[width=.75\textwidth,angle=0]{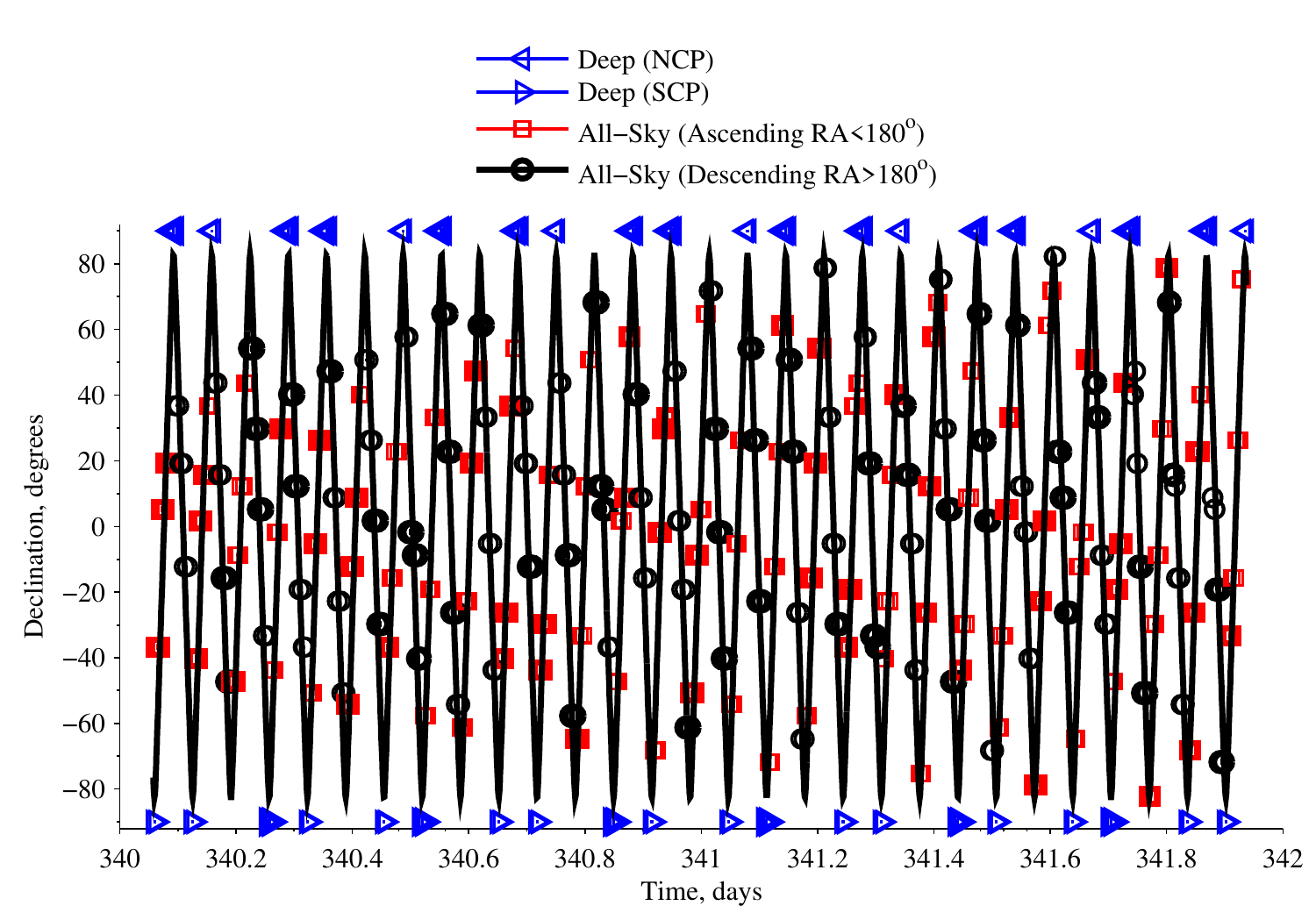}
   \label{fig:pointings_beta60}}
\end{center}
\caption{Representative schedules for a two day planning horizon.  
The black solid lines denote the orbit track on the celestial sphere, which establishes the available declinations as a function of time.
The symbols denote the survey pointings, which are chosen to satisfy the constraints, where each pointing is comprised of 4-9 steps where the step duration depends on the survey type.
There are large slews (90 sec) between successive pointings and small slews (10 sec) between successive steps (in a single pointing) not shown here.
%with $T_s=100$ sec, $T_d=140$ sec. 
\label{fig:rep_schedules}}
\end{figure}

\begin{figure}[!hb]
 \begin{center}
     \includegraphics[width=.6\textwidth,angle=0]{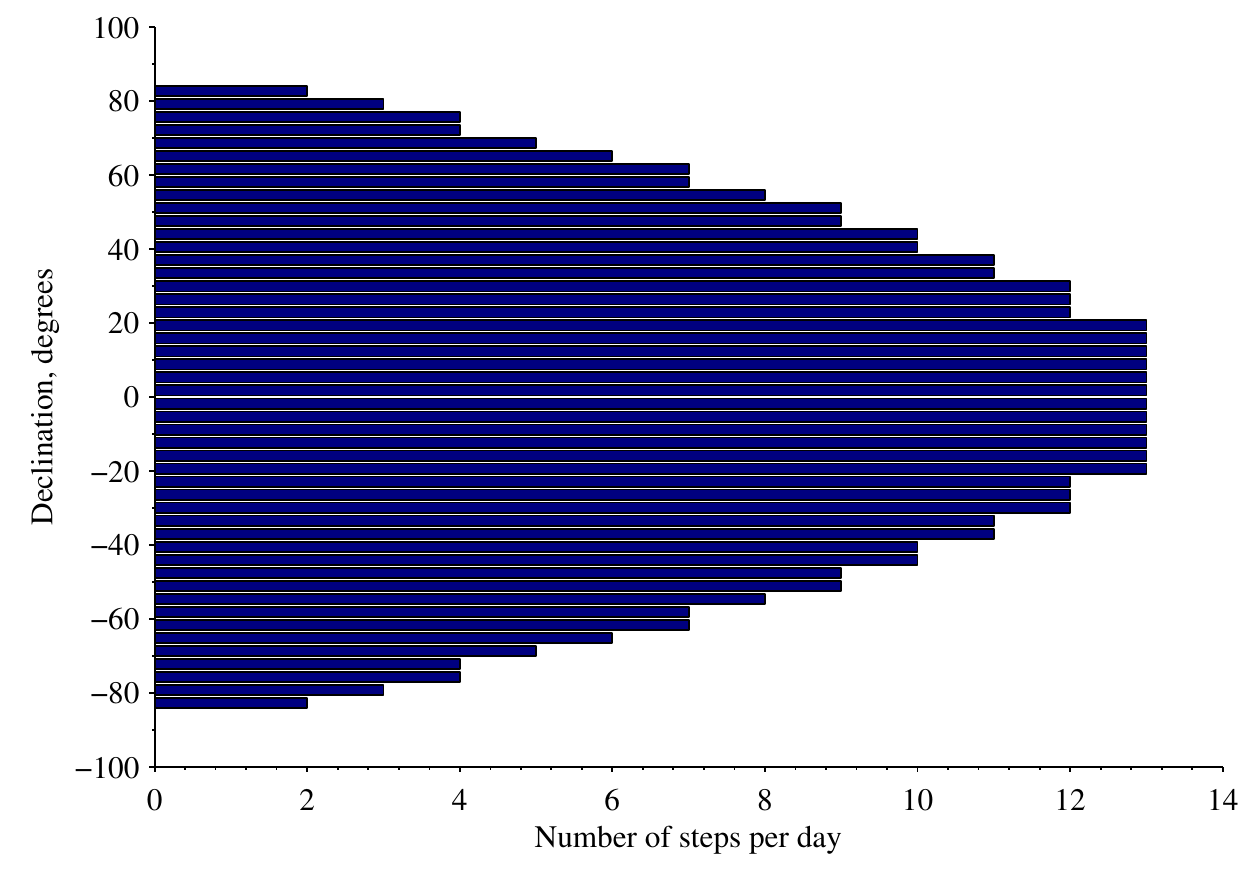}
 \end{center}
 \caption{Histogram of idealized number of steps per day to achieve redundancy requirements for all-sky survey.   This distribution in declination is readily implemented by the scenario illustrated in Fig. \ref{fig:rep_schedules}.
\label{fig:All_Sky_hist}}
\end{figure}

The algorithm described in section \ref{sec:obs_str}\ref{sec:sched_alg} is applied to the SPHEREx mission, where the targeted observations focus on the Deep and Galactic plane surveys.
Scheduling the NCP/SCP pointings (each with four steps) with a cadence of three quarters of orbits when $\beta=90^{\circ}$, and scheduling NCP/SCP pointings for two thirds the orbits when $\beta=60^{\circ}$) provided good science efficiency and enabled us to achieve the all-sky survey requirements.
%
%The galactic plane pointings  were scheduled for the appropriate fraction of orbits for the $\beta$ angle depending on the maximum allowable number of steps.
%
The survey is implemented in MATLAB\textregistered with orbital information generated with Systems Tool Kit (STK)\textregistered.
The algorithm was applied to realistic SPHEREx scenarios and the resulting schedules are shown in Figs. \ref{fig:rep_schedules_zoom}-\ref{fig:rep_schedules}.
The schedule is applied to two-day planning windows, and the steps fully cover two degrees in RA (as the orbit precesses at one degree per day).
The binned number of steps per day at each declination is shown in Fig. \ref{fig:All_Sky_hist}, which is essentially the steps collapsed from Fig. \ref{fig:rep_schedules}.
As the orbit precesses, a similar schedule is repeated at different RAs to cover the available celestial sphere. 
% for the two extreme $\beta$ angle cases.
%
%and detailed views are given in Figs. \ref{fig:rep_schedules_zoom}.
As expected, the results show 6 pointings/orbit for the $\beta=90^{\circ}$ case and 8 pointings/orbit for the $\beta=60^{\circ}$, which emerged naturally from the algorithm (i.e. it was not constrained).
The telecommunication operations (downloading and uploading from Earth ground stations) is not shown in Figs. \ref{fig:rep_schedules}-\ref{fig:rep_schedules_zoom} because it only occurs a few times a week, however appropriate time is allocated to these operations on average, see Table \ref{tab:schedule_beta90}.

Implementing the schedule for the $\beta=90^{\circ}$ case, as in Figs. \ref{fig:rep_schedules}-\ref{fig:rep_schedules_zoom}, results in a science efficiency of $78\%$.
%, which is a slight reduction in efficiency relative to the summary in Table \ref{tab:schedule_beta90}.
%
This science efficiency can be further improved to approach the optimized efficiency by using the generated schedules as initial guesses in a global optimization problem to maximize overall science time with the decision variable and model improvements discussed next.
The NCP/SCP pointing cadence,  placement in the schedule, and/or total number of pointings could also be decision variables, providing improved flexibility in overall scheduling towards improving efficiency.
The slew times could also be modeled more accurately based on actual slew angles for both short and long slews and the problem should be constrained to prevent any down time between science observations and slews, which will reduce overall slew times and improve efficiency.
%
%  improving the slew durations based on angular sizes of the slews, and/or  or by performing an overall optimization on the generated  integration durations, or repairing the generated schedule to ideally place the start and stop times
%
%
%\begin{figure}[!hb]
% \begin{center}
%     \includegraphics[width=1\textwidth,angle=0]{figs/multi_panel_shrunk}
% \end{center}
% \caption{Progressive coverage of all-sky survey at a single wavelength, where blue denotes a redundancy of half and green denotes a redundancy of one.  The approach achieves an overall efficiency of one in six months. \label{fig:rep_results}}
%\end{figure}

%\begin{figure}[!ht]
%\begin{center}
%%
%   \subfigure[3 months]{
%  \includegraphics[trim=4cm 4cm 4cm 4cm, clip=true, width=.45\textwidth,angle=0]{figs/all_sky_Raj/3months}
%   \label{fig:3months}}
%%
%   \subfigure[6 months]{+
%  \includegraphics[trim=4cm 4cm 4cm 4cm, clip=true,width=.45\textwidth,angle=0]{figs/all_sky_Raj/6months}
%   \label{fig:6months}}
%%
%   \subfigure[9 months]{
%  \includegraphics[trim=4cm 4cm 4cm 4cm, clip=true,width=.45\textwidth,angle=0]{figs/all_sky_Raj/9months}
%   \label{fig:9months}}
%%
%   \subfigure[12 months]{
%  \includegraphics[trim=4cm 4cm 4cm 4cm, clip=true,width=.45\textwidth,angle=0]{figs/all_sky_Raj/12months}
%   \label{fig:12months}}
%\end{center}
%\caption{Progressive coverage of all-sky survey at a single wavelength, where the strategy approach achieves an overall efficiency of one in six months. \label{fig:rep_results}}
%\end{figure}

\begin{figure}[!ht]
\begin{center}
   \subfigure[1 week]{
  \includegraphics[ width=.45\textwidth,angle=0]{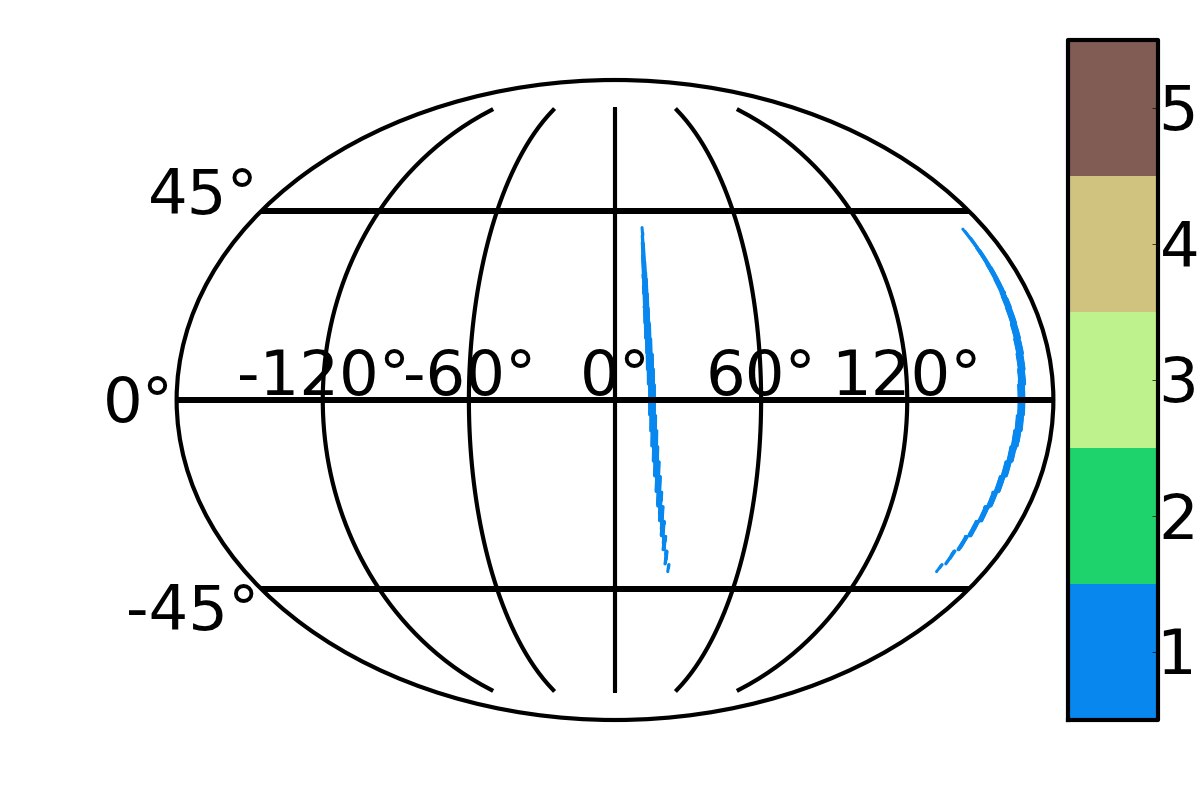}
   \label{fig:3months}}
   \subfigure[4 months]{
  \includegraphics[clip=true,width=.45\textwidth,angle=0]{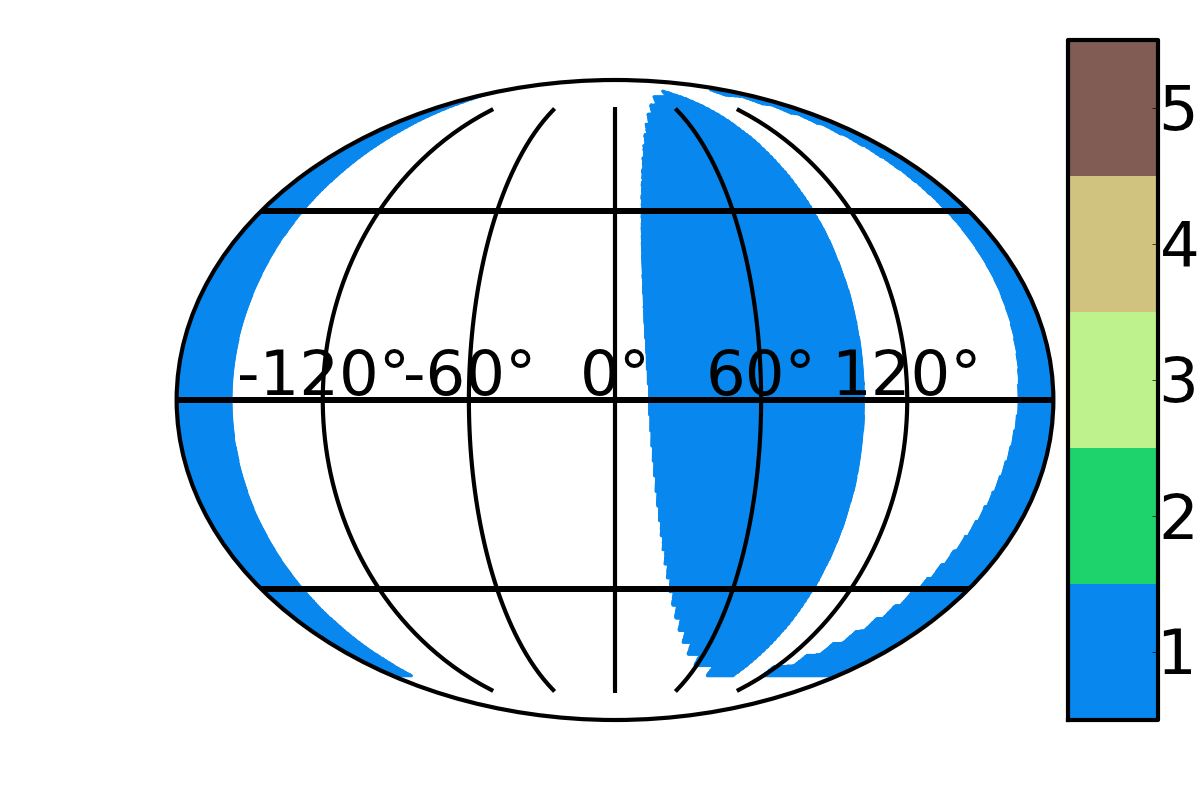}
   \label{fig:6months}}
   \subfigure[7 months]{
  \includegraphics[width=.45\textwidth,angle=0]{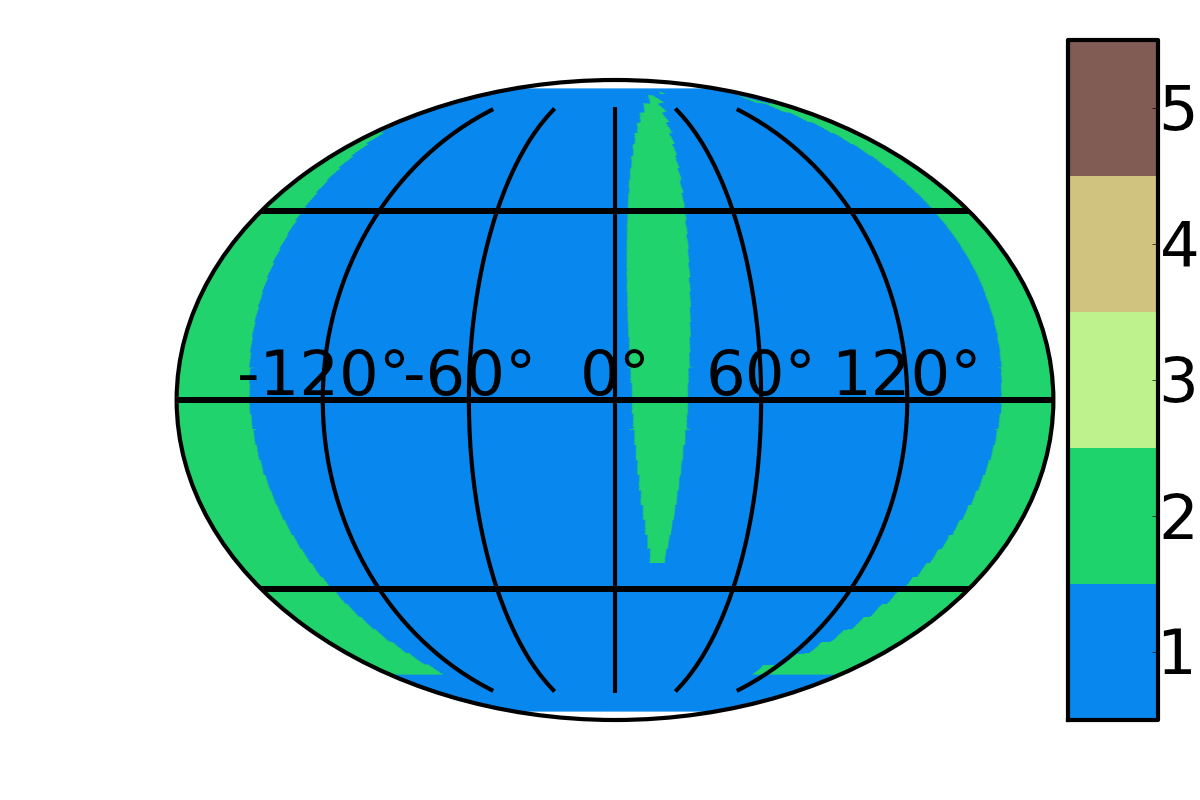}
   \label{fig:9months}}
   \subfigure[13 months]{
  \includegraphics[width=.45\textwidth,angle=0]{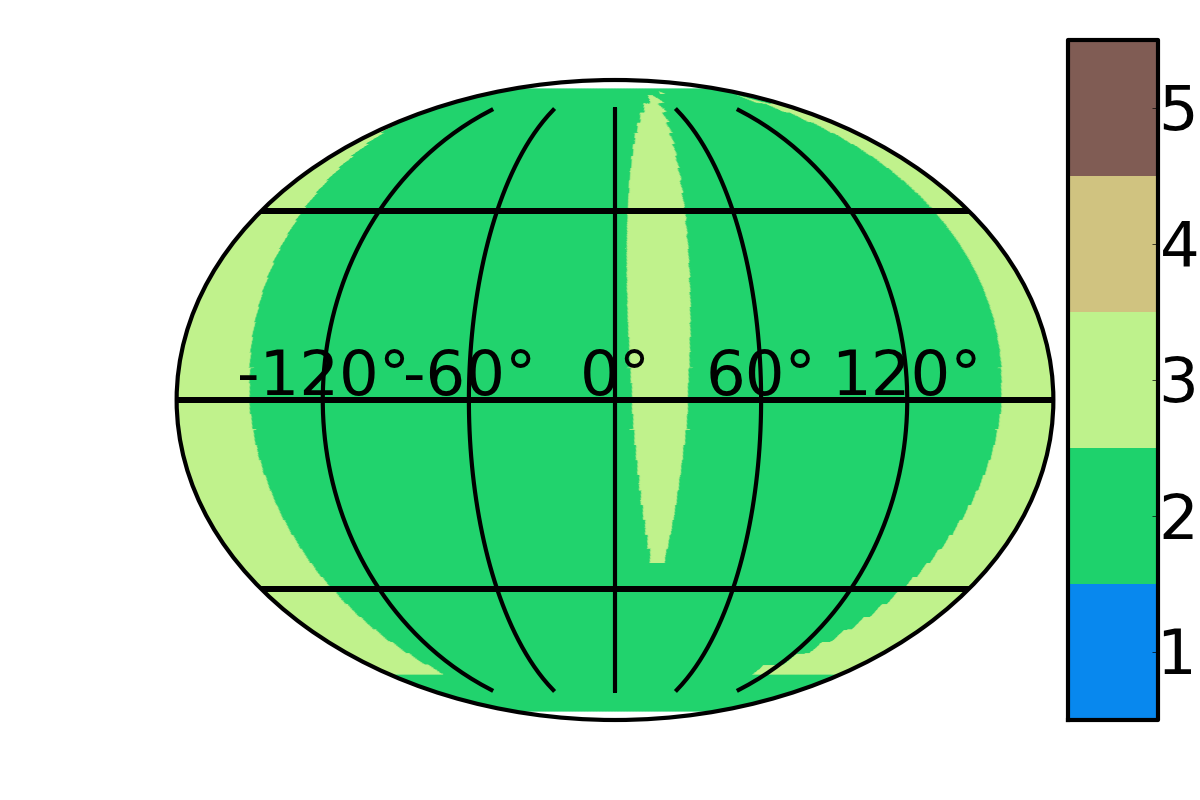}
   \label{fig:12months}}
\end{center}
\caption{Progressive coverage for all-sky survey, where the color shows the minimum wavelength coverage of every area of the sky.  The approach achieves an overall efficiency of one in six months. \label{fig:rep_results}}
\end{figure}

%Describe implementation, and NCP/SCP duration and cadence and Galactic plane duration and cadence for this orbit, probably in table form. (i.e. NCP/SCP two out of every three orbits when $\beta=90^o$, one out of every two orbits when $\beta=60^o$).
%

Representative coverage results of the all-sky survey over long-duration planning horizons are shown in Fig. \ref{fig:rep_results}.
These visualizations demonstrate that the observing scenario strategy achieves the required redundancy requirements for the all-sky survey, where uniform coverage with a redundancy of one in six months is achieved in the all-sky survey, and redundancy scales linearly with time.
%, as in Fig. \ref{fig:All_Sky_Example}.
%
%The deep survey achieves uniform and high redundancy over the area of interest $\geq83.5^{\circ}$, as in Fig. \ref{fig:Deep_Survey_results}, while the coverage surrounding this area shows the overhang effects which can contribute to the redundancy of the all-sky survey.

%
%\begin{figure}[!h]
%\begin{center}
%%
%   \subfigure[All-Sky Pointing Centers over six months to achieve a redundancy of one. 
%%   (Will actually be more of a hits map across entire sky, replace scale with 0-3 probably)
%   ]{
%  \includegraphics[width=.75\textwidth,angle=0]{figs/All_Sky_plotter} 
%   \label{fig:All_Sky_Example}}
%%
%   \subfigure[Deep Survey Coverage 
%   %(Will be updated with redundancy instead of number of total hits)
%   ]{
%  \includegraphics[width=.75\textwidth,angle=0]{figs/phi_365_uniform_fulllambda}
%   \label{fig:Deep_Survey_results}}
%%
%\end{center}
%\caption{Representative coverage results for the surveys 
%  \label{fig:rep_results}}
%\end{figure}

%\section{Additional Cases}
%\label{sec:add_cases}
%
%
%Discuss additional cases of galactic plane Survey and Moon avoidance.
%%
%Clean up the following discussion.

%
%%\subsection{Moon Avoidance}
%\section{Moon Avoidance}
%
%\label{sec:moon_avoid}
%\input{sections/moon_avoidance}

%\newpage
\section{Conclusion}
\label{sec:conclu}
This paper presented a general approach for scheduling all-sky surveys that accommodates targeted observations subject to dynamic constraints in the LEO environment.
The approach includes spacecraft and instrument constraints related to dynamic and interacting thermal and stray-light environments, which vary as a function of the orbit position relative to the Sun, Earth, and Moon.
% (due to the Earth, Sun, and Moon).
%, is dynamic throughout the year as the orbit precesses and the solar $\beta$ angle and location where it crosses the galactic plane evolves.
%
%We divide the global observing problem into several sub-problems and solve them in series, passing a reduced set of constraints between sub-problems.
%
The instrument and spacecraft configuration problem is solved to satisfy the Sun-avoidance criteria, which combined with the Earth-avoidance criteria leads to dynamic maximum observation time constraints throughout the year.
Targeted observations constrain the time available to schedule all-sky observations.
%the galactic plane and Sun drive when and the cadence of the Deep poles and galactic plane observations.
%
%These observations are input constraints to the 
A heuristic-based all-sky scheduling algorithm is presented that is designed to generate guaranteed-feasible solutions that achieve the all-sky redundancy requirements.
%
%The Moon-avoidance case is also addressed using an extension of the heuristic algorithm based on insights into constraint sensitivities.
%
The approach is applied to the proposed SPHEREx mission, which consists of a deep survey focusing on the celestial poles and an all-sky survey including a Galactic plane survey, where each survey has specific requirements and objectives.
Representative solutions are presented that achieve the survey goals for nominal and worst cases.
The scheduling efficiency of the all-sky survey is approximately $78 \%$, while idealized schedules are approximately $85 \%$ ($\beta=90^{\circ}$ case), where further optimization can be performed to approach the idealized efficiencies.
%

%

%This paper presents an approach to schedule multi-survey observing scenarios and how to effectively achieve the desired uniformity and redundancy coverage of galactic plane and all-sky surveys.
%
Beyond the general approach for efficiently accomplishing all-sky surveys, this paper presents insights for operating a spacecraft in the challenging LEO environment, where there may be Sun, Earth, and Moon avoidance constraints.
Furthermore, this approach could be utilized to map the Earth with  nadir-pointing spacecraft instead of the celestial sphere with a zenith-pointing spacecraft.
The general step-wise approach of solving several sub-problems presented in this paper has significantly reduced the number of constraints, variables, and objectives.
This approach is applicable to a larger class of spacecraft scheduling problems; however requires a solid understanding of the interactions of the problem  objectives and constraints.

%% file: lit_review_v2.tex
There is a large body of related research on spacecraft operations and scheduling.
Most of the scheduling approaches described in the literature involve a nadir-pointing spacecraft, however the formulations in the literature share similar dynamics, constraints, and objectives with the observing problem addressed in this paper.
%
%We review the approaches and algorithms in  well-studied problem of scheduling imaging spacecraft due to its extensive literature and similarities to the spacecraft communication scheduling problem.
%
%Second, 
We review the historic space-based all-sky survey observatories and scheduling work on pointed observations and discuss their similarities and differences relative to the problem addressed in this paper.

Several space-based observatory missions that performed or proposed to perform all-sky surveys.
%have had relatively simple observing strategies with limited constraints.
%
%We focus this review on survey-type missions
%
%For example, the TRACE SMEX mission launched in 1997 stared at the Sun for the first 7 months of its mission \cite{Handy}.  
%http://trace.lmsal.com/tag/
%
%The Kepler mission constantly stares at the same field of view in the sky where the constellations Cygnus and Lyra exist.
%, which was selected because it's an area rich in stars.
%http://kepler.nasa.gov/Mission/faq/#b1
%
The Infrared Astronomical Satellite (IRAS) was the first space-based observatory to perform a survey of the entire sky at infrared wavelengths from LEO, and it mapped 96$\%$ of the sky four times, each time at a different wavelength \cite{Emming}.
The IRAS satellite design and survey strategy were optimized to maximize the detection of point sources\footnote{http://www.ipac.caltech.edu/project/15}, and consisted of mapping out  "lunes" bounded by ecliptic meridians 30$^{\circ}$ apart\footnote{http://irsa.ipac.caltech.edu/IRASdocs}.
% and IRAS made significant discoveries of galaxies, dust disks, and the first images of the Milky Way Galaxy's core by focusing on cataloging fixed sources .
%
The Akari infrared astronomy satellite surveyed the entire sky in the near-, mid-, and far-infrared from LEO \cite{Jeong}.
This mission achieved continuous coverage by pointing in the zenith direction and scanning the sky continuously, while pointed observations are constrained by Sun and Earth constraints\footnote{http://www.ir.isas.jaxa.jp/ASTRO-F}.
Akari scanned $94\%$ of the sky twice and performed 5,000 pointed observations approximately 1.5 years following launch.
Wide-field Infrared Survey Explorer (WISE) performed an all-sky survey from LEO, imaging every part of the sky at least 8 times in ten months \cite{Edward}.
%, collecting images in the infrared-wavelength.
%
WISE's observing scenario consisted of continuously pointing in the zenith direction and taking a 47-arcminute field of view image every 11 seconds \footnote{http://wise.ssl.berkeley.edu/documents/FactSheet.2010.1.4.pdf}.
TBD- mention COBE, WMAP, and Planck.
The ROSAT All Sky Survey operated in LEO and consisted of 1378 distinct fields, where each scanning-mode observation covers 6.4 x 6.4$^{\circ}$ of sky \footnote{http://heasarc.gsfc.nasa.gov/docs/rosat/rass.html}.
The TESS mission will perform a a step and stare approach to monitor the full celestial sphere in a two-year mission by stepping the FOV 27$^{\circ}$ east every 27 days from a High Earth Orbit in 2:1 resonant orbit with the Moon \cite{Gangestad}.
Most of these observing strategies consisted of a step-and-stare approach with a zenith-pointed instrument to accomplish an all-sky survey, and most did not also consider pointed observations.

There has been considerable scheduling work focusing on pointed observations that is informative in the development of scheduling algorithms.
The Hubble Space Telescope (HST), launched in 1990, was one of the largest and most complex scheduling problems because 10,000 to 30,000 observations must be scheduled annually and operated in the challenging LEO environment and was subject to a large number of operational and scientific constraints \cite{Johnston_Spike}.
HST scheduling problems were formulated as constraint satisfaction problems and solved with search approaches that include multi-start stochastic repair strategies.
The James Webb Space Telescope (JWST), HST's predecessor to be launched in 2018, will operate from Earth-Sun L2 is also a complex multi-objective scheduling problem.
% with three objectives: to minimize schedule gaps, minimize the number of observations that miss their last scheduling opportunity, and minimize momentum build-up (caused by solar radiation pressure on its large sun sun-shield) \cite{Giuliano,Johnston_JWST_2008}.
%
JWST scheduling problems consist of three three objectives: to minimize schedule gaps, minimize the number of observations that miss their last scheduling opportunity, and minimize momentum build-up.
These scheduling problems are solved using evolutionary algorithms \cite{Giuliano,Johnston_JWST_2008}. 
%(caused by solar radiation pressure on its large sun sun-shield), 
%http://hubblesite.org/the_telescope/team_hubble/
%http://space.mit.edu/TESS/TESS/Mission_Overview.html
%
%Add other relevant missions and how they were scheduled.

%(TK: new sentence to highlight importance of this paragraph:) 

%We have not found a general approach to 
This paper develops a general approach to scheduling all-sky strategies that can accommodate diverse survey requirements and goals for the first time.
% does not exist in the literature.
%
Most observing problems in the literature consist of a zenith-pointing spacecraft that scans the sky quickly and it is not clear how they extend to accommodate additional surveys.
Furthermore, many are not restricted by dynamic thermal and stray-light constraints, nor do they consider instrument and spacecraft configuration decisions.
In general, the formulations and approaches in the literature (Refs. \cite{Emming,Jeong,Edward,Gangestad,Johnston_Spike,Giuliano,Johnston_JWST_2008}) have a different set of problem objectives, decisions, and constraints relative to the all-sky survey considered in this paper.
%
%For example, most of the \textit{EOS} literature ignores logistical constraints (e.g. thermal, stray-light) necessary in the SPHEREx problem.
%%
%In addition, SPHEREx has strict thermal and stray-light constraints, and is constantly performing science (i.e. pointing at a target or slewing between successive targets) to satisfy its three survey objectives and doesn't have the energy and data limitations modeled in some \textit{EOS} literature.
%towards 
%
%Most \textit{EOS} problems in the literature are not restricted by dynamic thermal and stray-light constraints, nor do they consider instrument and spacecraft configuration decisions, as in the SPHEREx observing problem.
%
%the SPHEREx observing problem.
%
%The observing problem is more complex than many of the observatory-type missions, including the TESS scheduling problem and the ROSAT problem as we have a significantly larger number of all-sky fields (150,000 per year) and must accommodate more constraints and additional surveys.
%
%[Update with more survey-type missions.]
% (i.e. TRACE and TESS).
%
%Furthermore, it differs from the complex HST and JSWT scheduling problems because the SPHEREx problem consists of over 150,000 observations per year and the three surveys must each achieve complete wavelength and spatial coverage over the planing horizon rather than having quantified priorities.
%h observation having a priority, all surveys must be completed throughout the planing horizon.
%
This paper develops a general approach for accomplishing an all-sky survey, which is flexible and can accommodate diverse objective goals and constraints.
Much of the observatory scheduling literature is informative in developing these models and algorithms.

%% file: SPHEREx_arXiv.bbl
\begin{thebibliography}{1}

\bibitem{Emming}
J.~G. {Emming}, R.~F. {Arentz}, C.~H. {Downey}, E.~C. {Long}, and L.~G.
  {Smeins}.
\newblock {Pulse circumvention circuit for the Infrared Astronomical Satellite
  telescope}.
\newblock In A.~{Boksenberg} and D.~L. {Crawford}, editors, {\em
  Instrumentation in astronomy V}, volume 445 of {\em Society of Photo-Optical
  Instrumentation Engineers (SPIE) Conference Series}, pages 254--263, January
  1984.

\bibitem{Jeong}
Woong-Seob Jeong, Soojong Pak, Hyung~Mok Lee, Takao Nakagawa, Minjin Kim,
  Sang~Hoon Oh, Hidehiro Kaneda, Sin’itirou Makiuti, Mai Shirahata, Shuji
  Matsuura, Mikhail~A Patrashin, Chris Pearson, and Hiroshi Shibai.
\newblock {ASTRO-F/FIS} observing simulation including detector
  characteristics.
\newblock {\em Advances in Space Research}, 34(3):573 -- 577, 2004.
\newblock Astronomy at IR/Submm and the Microwave Background.

\bibitem{Edward}
Edward~Wright et~al.
\newblock The wide-field infrared survey explorer (wise): Mission description
  and initial on-orbit performance.
\newblock {\em The Astronomical Journal}, 140:1868, 2010.

\bibitem{Gangestad}
Joseph~W Gangestad, Gregory~A Henning, Randy~R Persinger, and George~R Ricker.
\newblock A high earth, lunar resonant orbit for lower cost space science
  missions.
\newblock Jun 2013.
\newblock Comments: 15 pages, 15 figures, to be presented at AAS/AIAA
  Astrodynamics Specialist Conference, August 2013.

\bibitem{Johnston_Spike}
M.~D. Johnston, ed. M.~Zweben G.~E.~Miller, and Morgan~Kaufmann M.~Fox.
  San~Mateo.
\newblock Spike: Intelligent scheduling of hubble space telescope observations.
\newblock {\em Intelligent Scheduling}, pages 391--422, 1994.

\bibitem{Giuliano}
M.~Giuliano and M.~D. Johnston.
\newblock Multi-objective evolutionary algorithms for scheduling the james webb
  space telescope.
\newblock In {\em International Conference on Automated Planning and Scheduling
  (ICAPS)}, Sydney, Australia, 2008.

\bibitem{Johnston_JWST_2008}
Mark D.~Johnston Mark E.~Giuliano.
\newblock Multi-objective evolutionary algorithms for scheduling the james webb
  space telescope.
\newblock {\em Eighteenth International Conference on Automated Planning and
  Scheduling (ICAPS 2008)}, 2008.

\bibitem{Rosenberg}
K.~Rosenberg, K.~Hendrix, D.~Jennings, D.~Reuter, M.~Jhabvala, and A.~La.
\newblock Logarithmically variable infrared etalon filters.
\newblock {\em SPIE Proc. Opt. Thin Films IV N. Dev.}, 2262:25--27, 1994.

\end{thebibliography}
